\documentclass[%
 reprint,
 amsmath,amssymb,
 aps,prl,
floatfix,
]{revtex4-2}

\usepackage{graphicx}
\usepackage{dcolumn}
\usepackage{bm}

\usepackage{graphicx,amsmath,amsfonts,amscd,amsthm,amssymb,pstricks}
\usepackage{xcolor}
\usepackage{dcolumn}
\usepackage{bm}
\usepackage{mathrsfs}
\usepackage{textcomp}
\usepackage{esvect}
\usepackage{float}
\usepackage[USenglish,british]{babel}
\usepackage{environ}
\usepackage{xifthen}
\usepackage{xargs}		
\usepackage{hyperref}
\usepackage{physics,mathtools}
\usepackage[separate-uncertainty = true]{siunitx}
\usepackage{multirow}
\usepackage{comment}
\usepackage{enumerate}
\usepackage{tikz}	
\usetikzlibrary{arrows, calc, matrix, patterns, decorations.markings, shapes, decorations.pathmorphing, shadows.blur}
\usepackage[utf8]{inputenc}
\usepackage{enumitem}
\usepackage{caption,booktabs}
\usepackage{subcaption}
\begin{document}

\title{Proof-of-principle experiment of a novel beam extraction over millions of turns using stable resonance islands and bent crystal}

\author{D. E. Veres}
\altaffiliation[Also at ]{Goethe University, 60323 Frankfurt am Main, Germany}
\affiliation{CERN, Esplanade des Particules 1, 1121 Meyrin, Switzerland}

\author{P. Arrutia}
\affiliation{CERN, Esplanade des Particules 1, 1121 Meyrin, Switzerland}

\author{S. Cettour Cave}
\affiliation{CERN, Esplanade des Particules 1, 1121 Meyrin, Switzerland}

\author{B. H. F. Lindstr\"om}
\altaffiliation[Also at ]{John Adams Institute at Royal Holloway, University of London, Egham, United Kingdom}
\affiliation{CERN, Esplanade des Particules 1, 1121 Meyrin, Switzerland}

\author{M.~Giovannozzi}
\email{massimo.giovannozzi@cern.ch}
\affiliation{CERN, Esplanade des Particules 1, 1121 Meyrin, Switzerland}

\author{L. E. Pauwels}
\altaffiliation[Also at ]{Universit\'{e} Libre de Bruxelles, Avenue F.~D.~Roosevelt 50, B-1050, Brussels, Belgium}
\affiliation{CERN, Esplanade des Particules 1, 1121 Meyrin, Switzerland}

\author{F. F. Van der Veken}
\affiliation{CERN, Esplanade des Particules 1, 1121 Meyrin, Switzerland}

\author{F. M. Velotti}
\affiliation{CERN, Esplanade des Particules 1, 1121 Meyrin, Switzerland}

\author{G. Franchetti}
\altaffiliation[Also at ]{Goethe University, 60323 Frankfurt am Main, Germany}
\affiliation{GSI Helmholtzzentrum f\"ur Schwerionenforschung GmbH, Planckstraße 1, 64291 Darmstadt, Germany}

\date{\today}

\begin{abstract}
A recent study~\cite{DEV_sx_PRR} has introduced an advanced method aimed at extracting from a circular particle accelerator over millions of turns using stable islands and a bent crystal. This technique leverages the strength of non-linear beam dynamics along with adiabatic trapping and transport within stable islands, in combination with a bent silicon crystal, to provide an efficient extraction of hadron beams over millions of turns. The positive and encouraging results of the comprehensive numerical simulations were validated through a successful proof-of-principle experiment at the CERN Super Proton Synchrotron, which demonstrated the feasibility of the technique. This paper presents a detailed discussion and analysis of the experimental results. 
\end{abstract}

\maketitle

Nonlinear beam dynamics has enabled new strategies for controlling and manipulating charged-particle beams in circular accelerators. 
A prominent example is Multi-Turn Extraction (MTE)~\cite{PhysRevLett.88.104801,PhysRevSTAB.7.024001}, which was developed at the CERN Proton Synchrotron (PS) and is now the standard extraction scheme for delivering high-intensity fixed-target proton beams to the Super Proton Synchrotron (SPS)~\cite{Borburgh:2137954,PhysRevAccelBeams.20.061001,PhysRevAccelBeams.22.104002,PhysRevAccelBeams.25.050101}.

Similarly, bent crystals have transformed accelerator physics, as they offer a means to steer particle trajectories~\cite{lindhard:1965}. 
Following extensive studies~\cite{carrigan:1990,elsener:1996,biryukov:1997,PhysRevLett.87.094802,PhysRevSTAB.6.033502,fliller:2005,PhysRevLett.98.154801,PhysRevSTAB.9.013501,PhysRevSTAB.11.063501,PhysRevLett.102.084801,scandale:2010,scandale:2013,PhysRevAccelBeams.21.014702,shiltsev:2019,PhysRevApplied.14.064066,scandale:2022}, bent crystals have become a key element of the operational collimation system at the CERN Large Hadron Collider (LHC)~\cite{scandale:2016,ion-beam-channelling,PhysRevAccelBeams.27.011002,PhysRevAccelBeams.28.051001}. 
In addition, the possibility of using crystals to extract the beam halo for fixed-target experiments at the LHC~\cite{fomin:2017,bagli:2017,botella:2017,redaelli:ipac18-tupaf045,PhysRevLett.123.011801,fomin:2019,mirarchi:2020,Dewhurst:2023cth} is currently under investigation within the Physics Beyond Colliders programme at CERN~\cite{Barschel:2653780,jaeckel:2020,pbc}.

Building on these lines of research, a new beam–manipulation scheme has recently been proposed that combines stable islands, generated by sextupoles and octupoles, with bent crystals to realize an efficient method for extracting charged particles over millions of turns~\cite{DEV_sx_PRR,DEV_sx_NIMA}. 
This innovative scheme overcomes several drawbacks of the conventional method, which relies on the third-order unstable resonance driven by sextupoles. 
In the standard approach, the unbounded branches of the separatrix send particles to large amplitudes until they intercept the gap of an extraction septum (see, e.g., Refs.~\cite{Tuck:1951,LeCouteur:1951,gordon:1958,hammer:1961,kobayashi:1967,gordon:1971,PIMMSvol1,Hardt:1025914,PulliaSX} and references therein). 
The motion along the separatrices is inherently chaotic, making interactions between the particles and the septum foil unavoidable. 
These interactions lead to irradiation that can shorten the operational lifetime of accelerator components, complicate maintenance~\cite{Fraser2017JACoWS, PhysRevAccelBeams.22.123501}, and restrict the maximum extractable beam intensity. 
Moreover, the horizontal profile of the extracted beam is often difficult to match to the optics of the downstream transfer line.

The use of octupoles has been considered (see~\cite{PhysRevAccelBeams.22.043501,PhysRevAccelBeams.22.123501} and references therein), and several implementations of crystal-assisted slow extraction have already been demonstrated and used in operation, including non-resonant extraction of the beam halo~\cite{Fraser:IPAC2017-MOPIK048,giacomelli:ipac25-tupb027} and shadowing of the electrostatic septum with bent crystals~\cite{velotti:ipac17-mopik050,esposito:ipac19-wepmp028,velotti:ipac19-thxxplm2,PhysRevAccelBeams.22.093502,velotti:ipac22-wepost013} to reduce beam losses during extraction. However, the new scheme differs fundamentally as it relies on stable islands to confine particles and drive them to large amplitudes, where they eventually encounter a bent silicon crystal. The crystal then steers these particles away, effectively separating them all from the remaining circulating beam. The dynamics both within and outside the stable islands is entirely regular, allowing multi-turn effects to be harnessed, which in turn substantially improves the channeling efficiency and hence the extraction efficiency.

The promising results of the numerical simulations motivated experimental studies to evaluate how the proposed beam manipulation actually behaves in an operational accelerator. 
The SPS provides a suitable environment for a proof-of-principle beam experiment, and its implementation has been divided into two stages: a first stage focused on generating and probing the desired phase-space topology and a second stage dedicated to investigating the complete beam manipulation process, including beam trapping, transport, and channeling.

\begin{figure*}[htb]
    \centering
    \includegraphics[width=\linewidth]{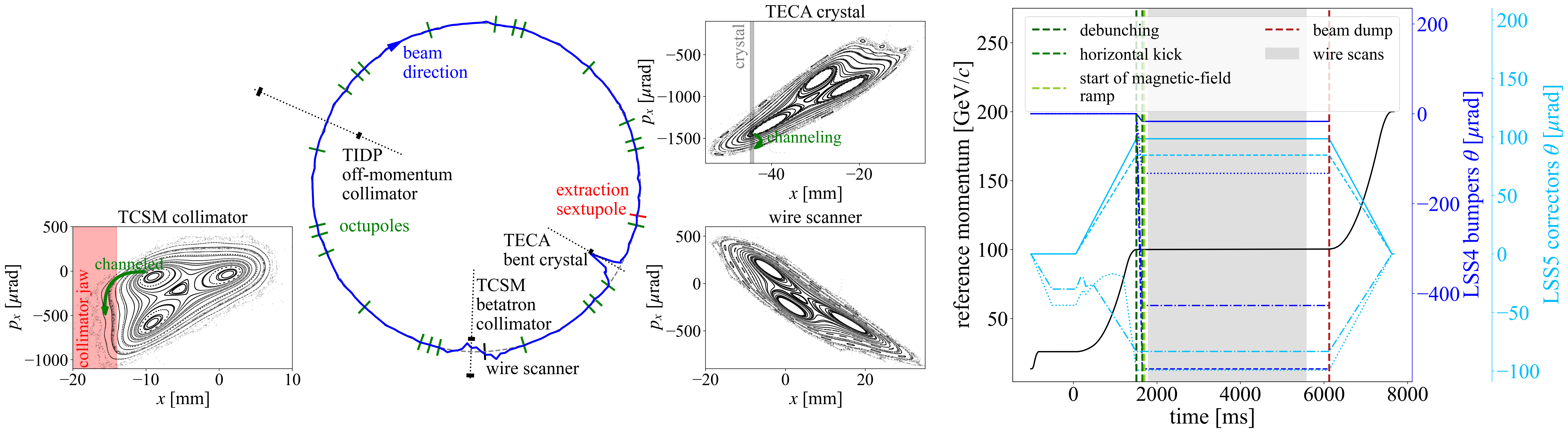}
    \caption{Left: Illustration of key devices used in the proof-of-principle measurement at the SPS. Collimators' jaws and crystal positions are shown in black, the beam orbit is shown in blue. Non-linear elements (octupoles and extraction sextupole) are marked with green and red solid lines, respectively. Magnitudes are not to scale for better visibility. The phase-space portraits at the crystal and betatron collimator are also shown. Right: Magnetic cycle (black), and deflection angle of magnets used for orbit bumps in LSS4 (blue) and LSS5 (light blue) as function of time. Key events are marked with colored dashed vertical lines.}
    \label{fig:ring_with_collimators-cycle_info}
\end{figure*}

The first stage was successfully achieved and the phase-space topology with stable islands of the third-order resonance was reconstructed, and beam splitting, due to trapping in the islands, and transport were demonstrated for the first time at the SPS, and found to be in excellent agreement with numerical simulations~\cite{DEV_IPAC25_SPS_LPR}. 
Hence, a dedicated proof-of-principle measurement of the novel extraction with stable islands and bent crystal was conducted in the same accelerator. 
Given the fragile nature and importance of the SPS electrostatic septa used for slow extraction, the test had to be carried out without attempting beam extraction to avoid any accidental damage to the devices. 
Therefore, the measurement was designed to demonstrate controlled particle transport by means of the stable islands up to the amplitude of the bent silicon crystal (TECA) in the Long Straight Section (LSS) 4 of the SPS. 
The particles in the stable islands can then be efficiently channeled by the crystal, which provides a deflection that separates the particles that interacted with the crystal from those that did not, driving the first onto the betatron collimator (TCSM) in LSS5. 
Therefore, monitoring beam losses at the collimator amounts to demonstrating that the core principle of the proposed extraction scheme functions properly and that the beam will ultimately be extracted.
The position of the devices relevant for the experimental test, the orbit of the beam around the SPS ring, including special orbit bumps required for the measurement, as well as the horizontal phase-space portraits at the location of the crystal and collimator are shown in Fig.~\ref{fig:ring_with_collimators-cycle_info} (left).

As extracting the beam was not possible, the measurement was conducted at \SI{100}{\giga\eV} instead of \SI{400}{\giga\eV}, the typical energy of slow-extraction beams, and using the same ring optics as for LHC-type beams, similar to the case of preliminary measurements~\cite{DEV_IPAC25_SPS_LPR}, which allowed easy transfer of already established settings. 
Note that the main drawback of this choice is that the phase advance between the crystal and the collimator is such that the separation between channeled particles and those circulating in the islands is largest at the collimator on the second turn after channeling. 
Operationally, this configuration would not be ideal, as some channeled particles may interact with the crystal a second time amorphously before being intercepted by the collimator (or extracted). 
However, the conditions were acceptable for a proof-of-principle test.

The magnetic cycle used for the measurement was \SI{9.6}{\second} long with an approximately \SI{4}{\second}-long plateau at \SI{100}{\giga\eV} where the beam manipulation was performed, as shown in Fig.~\ref{fig:ring_with_collimators-cycle_info} (right). 
During the acceleration from injection energy of \SI{26}{\giga\eV} to \SI{100}{\giga\eV}, the tunes were moved close to the third-order resonance ($Q_x=26.313$, $Q_y=26.11$), and the strength of the non-linear elements used to generate the stable islands was increased to nominal strength. 
A single extraction sextupole ($k_2=\SI{0.399}{m^{-3}}$~\footnote{Note that the normalized nonlinear gradient is defined as $k_n=1/(B \rho) \dd^n B_y/\dd x^n$, where $B \rho$ is the rigidity of the beam.}) was used to drive the resonance and control the orientation of the islands, and all 24 octupoles (LOF) located in the arcs where the horizontal beta-function features a maximum were powered ($k_3=\SI{-6.0}{m^{-4}}$) to ensure the generation of the islands.

Strong orbit correctors in LSS4 were also powered during acceleration to generate a closed-orbit bump at the crystal (see Fig.~\ref{fig:ring_with_collimators-cycle_info} (right)), whose closest distance to the beam is limited to \SI{-39}{\milli\meter} with respect to the center of the beam pipe, and was positioned at \SI{-44}{\milli\meter} during the measurement.

To accommodate for the larger beam size at lower energy, the collimator jaw was kept at an intermediate amplitude and orbit correctors in LSS5 were powered up during acceleration to generate a closed-orbit bump towards the negative (right) jaw during beam manipulation as shown in Fig.~\ref{fig:ring_with_collimators-cycle_info}. 

The same approach was used to drive the trapping and transport in stable islands~\cite{DEV_sx_NIMA}, i.e. the Constant Optics Slow Extraction (COSE)~\cite{sx_COSE}, in which the fields of all magnets in the SPS ring were ramped slightly after turning off the RF cavities, thus inducing debunching of the beam, so that the beam gradually gained a coherent off-momentum with respect to the ring optics. 
Combined with the large negative chromaticity ($Q_x'=-21.5$) this steadily pushes the particles' tune across the third-order resonance.

The test was carried out with four low-intensity bunches of $10^{10}$ protons with normalized emittances of approximately \SI{0.56}{\micro\meter} in each transverse plane. 
The small horizontal emittance allowed for applying a horizontal kick to the beam before the start of manipulation to generate a hollow distribution that helps ensure the full depletion of the beam intensity during the entire process.

The beam measurement was conceived to show that the entire beam intensity can be depleted via channeling in the crystal and subsequent interaction with the collimator, without inducing losses elsewhere in the ring. This objective was achieved by implementing three distinct steps that together form the proof-of-principle experiment: $i)$ Perform a scan of the position of the collimator jaw with the crystal fully retracted, to determine how the onset of beam–collimator interception depends on the position of the jaw. As the jaw is moved farther from the beam center, the stable islands require progressively more time to grow to the corresponding amplitude; $ii)$ Repeat the same jaw-position scan with the crystal set in amorphous orientation; $iii)$ Repeat the procedure with the crystal oriented for channeling. The latter two measurement sequences make it possible to isolate the effect of the crystal, demonstrating that the beam interaction with the collimator jaw becomes insensitive to the exact jaw position. For both amorphous and channeling configurations, the crystal is sufficiently transparent that dominant beam losses are expected to occur at the collimator. However, the local losses at the crystal itself were predicted to be significantly lower in the channeling orientation than in the amorphous one. 

Together, these measurements enable a comprehensive characterization of the three configurations in the proof-of-principle test. In addition to confirming the expected locations and magnitudes of the beam losses, it was essential to observe the formation of stable islands in phase space and a smooth, continuous decrease in beam intensity. This behavior confirms that particles are transported to large amplitudes within the islands and that the process can indeed be regulated to avoid abrupt loss events.


\begin{figure}
    \centering
    \includegraphics[width=0.7\linewidth]{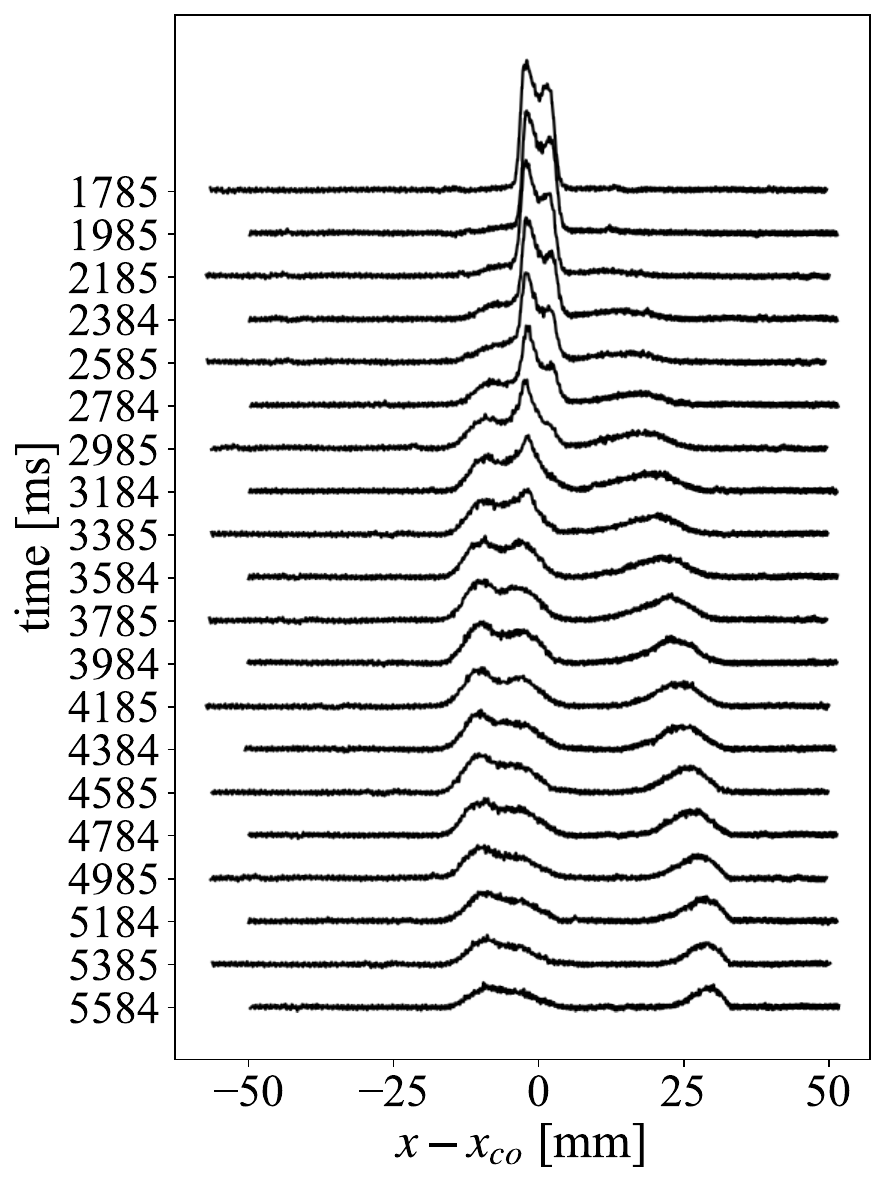}
    \caption{Horizontal beam profile measured using a wire scanner throughout the proof-of-principle measurement. Scans were taken in different machine cycles, and the closed orbit was subtracted from the wire position.}
    \label{fig:wirescans}
\end{figure}

To confirm the presence of stable islands and of the beam trapping and transport, beam profile measurements were taken with a wire scanner at different times during manipulation. 
For these measurements, both the collimator and the crystal were retracted to allow the islands to grow to large amplitudes and separate well for better visibility. 
The evolution of the horizontal beam profile from the hollow distribution to one that shows three peaks corresponding to the beam trapped in the islands is shown in Fig.~\ref{fig:wirescans}. 
Unfortunately, with the optimal orientation of the islands at the crystal location, two islands overlap slightly at the wire scanner location (see Fig.~\ref{fig:ring_with_collimators-cycle_info}, left), nevertheless, the islands are clearly visible. 
At around \SI{4000}{\milli\second} in the cycle, the islands reached the amplitude of the off-momentum collimator (TIDP) and the overall intensity started to deplete, thus reducing the signal from the final measured profiles.

\begin{figure}
    \centering
    \includegraphics[width=\linewidth]{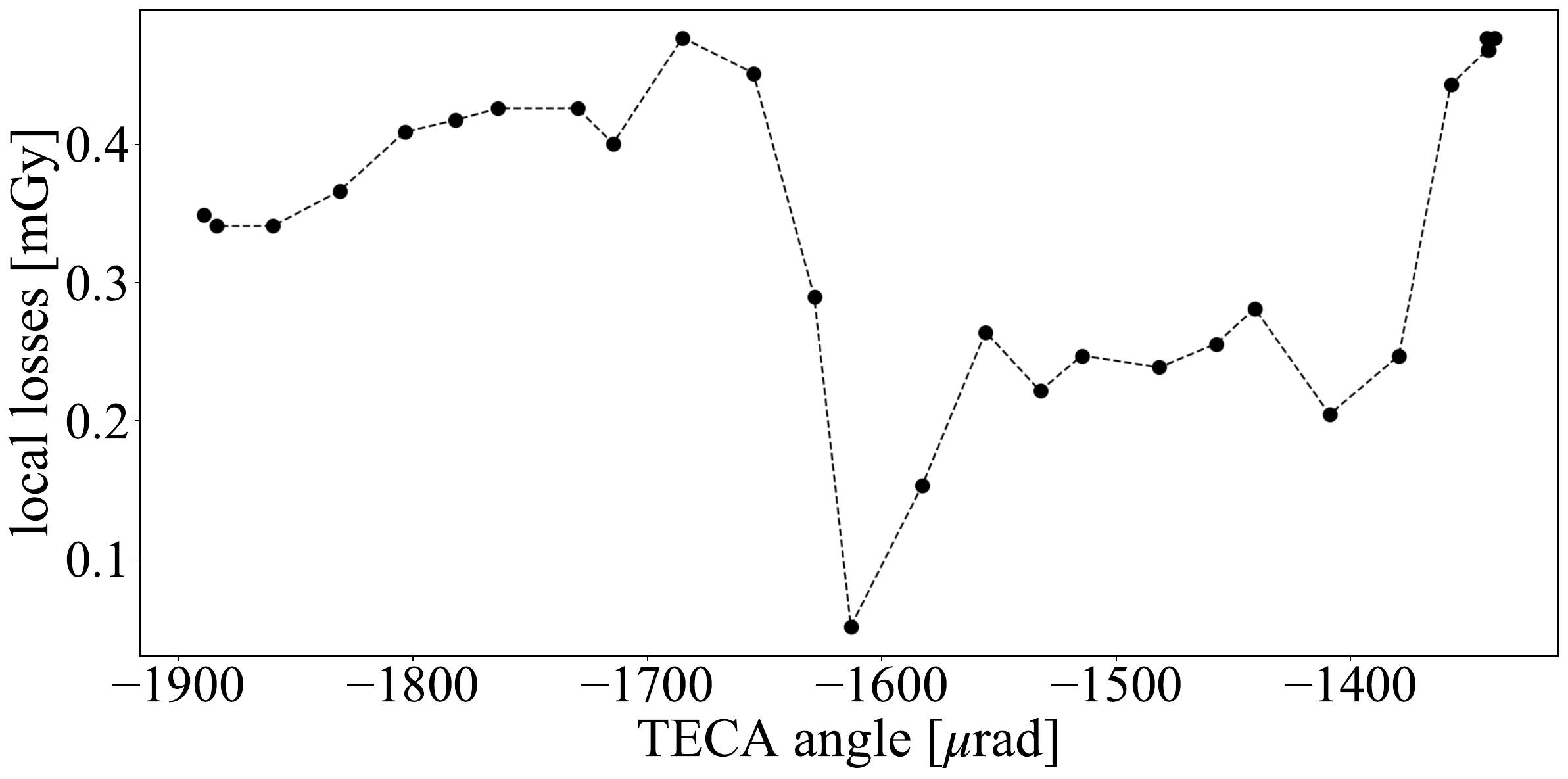}
    \caption{Local losses at the location of the bent crystal as a function of the crystal orientation. A minimum in losses, corresponding to channeling, is clearly visible.}
    \label{fig:teca_alignment}
\end{figure}

An integral part of the experimental setup is the determination of the crystal angle that provides beam channeling. 
Figure~\ref{fig:teca_alignment} shows local losses at the crystal location as a function of the crystal angle. 
Channeling was achieved at an angle around \SI{-1600}{\micro\radian} where the local losses decreased to a minimum. 
The amorphous region and volume reflection plateau can also clearly be identified to the left and right of the channeling well, respectively.

\begin{figure}
    \centering
    \includegraphics[width=\linewidth]{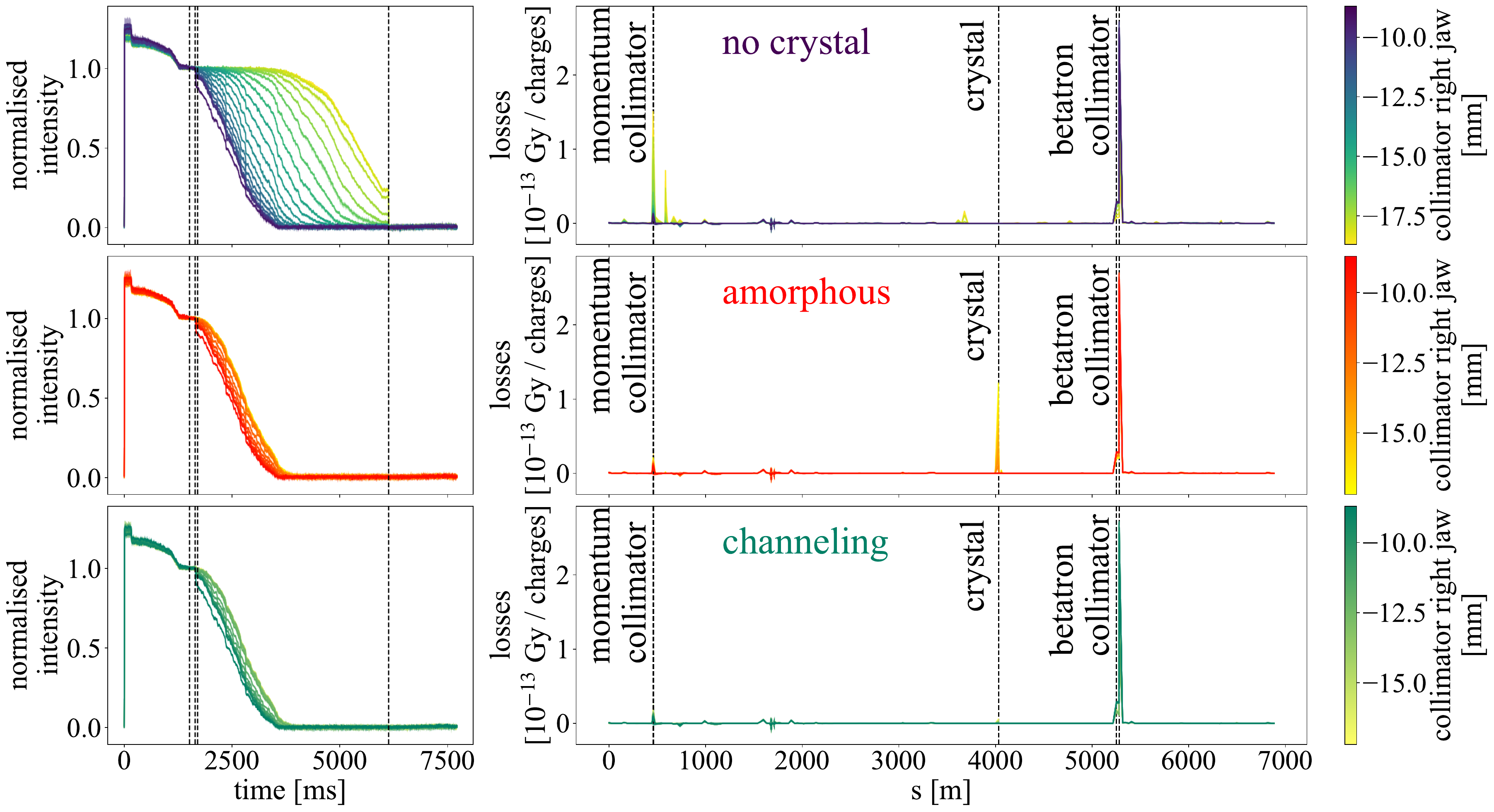}
    \caption{Intensity evolution (left) and total losses around the ring (right) for all collimator jaw positions with different crystal configurations. The debunching, horizontal kick, start of the magnetic-field ramp, and the dump are marked by vertical black dashed lines on the panels on the left. The location of off-momentum collimator, crystal and collimator BLMs are marked by vertical black dashed lines on the panels on the right.}
    \label{fig:measurement_overview}
\end{figure}

The key results of the proof-of-principle test are shown in Fig.~\ref{fig:measurement_overview}. 
There, the evolution of the total beam intensity recorded by the Beam Current Transformer (BCT) as well as the total losses recorded by Beam Loss Monitors (BLM) along the SPS ring as a function of collimator jaw position with the crystal in different orientations are shown. The BCT data are normalized to 1 at the beginning of the intermediate flat top. 
Intensity depletion is steady as desired, and the dependence of the timing of the beam-collimator interaction on the collimator jaw position is apparent when the crystal is retracted. However, such a dependence disappears when the crystal is inserted, as expected, and the ring location of the beam losses matches the expectations nicely. 
The majority of the beam is intercepted by the collimator, except when the jaw is retracted beyond the off-momentum collimator, and sizable losses at the crystal occur only when it is in amorphous orientation. This confirms that the beam intensity can indeed be efficiently depleted using the novel approach with minimal undesired losses. 
Furthermore, the ring location of the losses agrees with what is expected on the basis of the processes in action, namely, beam trapping and transport in stable islands and interaction with the bent silicon crystal.

\begin{figure}
    \centering
    \includegraphics[width=\linewidth]{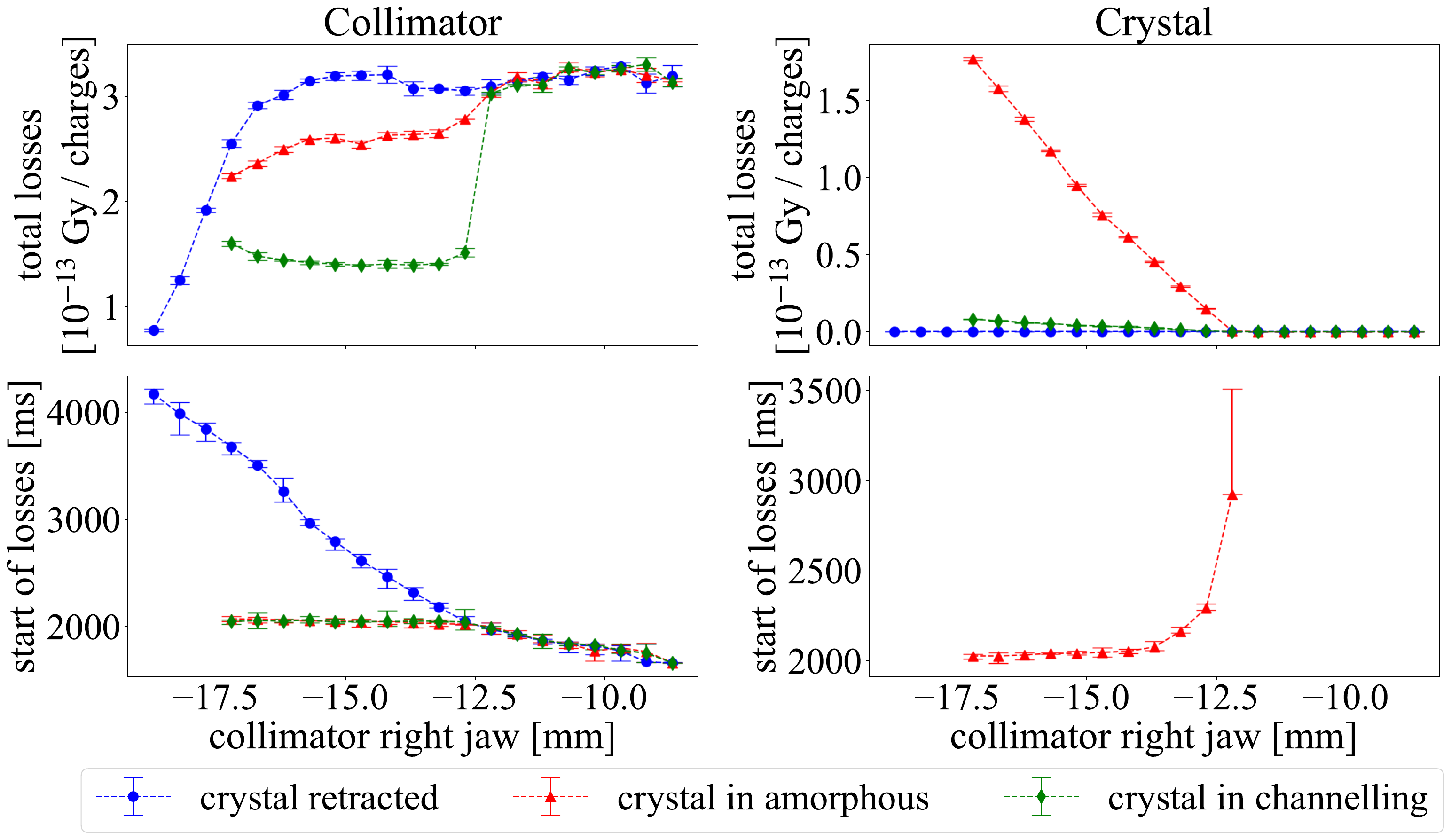}
    \caption{Total losses (top row) and start of losses (bottom row) at the collimator (first column) and crystal (second column) as a function of collimator jaw position for the three crystal configurations (color).}
    \label{fig:measurement_summary}
\end{figure}

The summary of total losses and start of losses at the crystal and collimator as a function of the jaw position of the collimator for all crystal configurations is shown in Fig.~\ref{fig:measurement_summary}. 
The behavior of the losses at the crystal is clear: when the device is retracted or in channeling orientation, the losses are essentially the same and independent of the collimator jaw position. 
However, a clear dependence is seen for the case of amorphous orientation of the crystal. 
In this configuration, the losses decrease as the collimator jaw approaches the beam center, making the collimator the global aperture limit of the ring. 
At this stage, the losses at the collimator are also independent of the crystal configuration, which is according to expectations. 
When the collimator jaw is further away from the center of the beam, the three configurations behave differently. The amorphous configuration has intermediate losses as part of the beam is lost at the crystal (see Fig.~\ref{fig:measurement_overview}, right panels). 
When in channeling orientation, losses show a large drop. 
In this configuration, the channeled particles hit the collimator jaw with an impact parameter up to \SI{3.5}{\milli \meter}, resulting in a smaller shower detected by the BLMs some \SI{30}{\meter} downstream than in the other two configurations where the jaw was simply scraping the beam. 
This is a clear signature of coherent trapping, transport, and channeling processes in action to separate a part of the beam from the circulating one. 

The start of losses was taken to be the time at which the cumulative losses at the respective BLMs reached 5\% of the total losses registered by the BLMs. 
The dependence of the start of losses on the collimator jaw position at the collimator features two regimes: when the collimator becomes the global aperture bottleneck, no difference is found for the three configurations. 
However, when the jaw position is farther away, an almost linear dependence is shown for the case with the crystal retracted, whereas no dependence is observed for the other two crystal configurations. All this is in full agreement with expectations. 

The onset of losses at the crystal location warrants particular attention because only in the amorphous configuration are the losses sufficiently large to yield meaningful information about when they begin. This is evident in the plot: there is no dependence on the collimator jaw position until the losses reach a sufficiently low level, at which point a rapid increase in the inferred starting time is observed. This steep variation is an artifact caused by the combination of losses that are initially too small and the finite BLM resolution, which also explains why the other two crystal configurations are not included.

In this Letter, we presented the design and results of a successful proof-of-principle experiment of the new beam manipulation scheme. A proton beam was captured in stable islands, and the particles were driven to large amplitudes, where they interacted with a bent crystal. Through channeling in the crystal, these protons were deflected and detected downstream. This procedure produced a continuous and smooth decrease in beam intensity, with beam losses confined exclusively to the collimator employed to monitor the process, mimicking an extraction process. All measured physical quantities consistently support the principles of the proposed extraction method, representing a highly promising outcome that paves the way for considering this approach in future circular accelerator applications, following thorough implementation studies.

\begin{acknowledgments}
The authors thank the colleagues from the CERN Accelerators and Beam Physics group, in particular H.~Bartosik and S.~Redaelli, for their expert support and advice on the use of the collimator as diagnostic tool, as well as the experts from the Accelerator Beam Transfer group for discussions and the SPS Operations crew for constant support during the experimental studies, and F.~Asvesta, P.~Bestmann, K.~S.~B.~Li, G.~Papotti and B.~Salvant for excellent support during the various stages of the experimental activities.
\end{acknowledgments}

%
\bibliography{mybib,mybib2}

\providecommand{\noopsort}[1]{}\providecommand{\singleletter}[1]{#1}%
\begin{thebibliography}{62}%
\makeatletter
\providecommand \@ifxundefined [1]{%
 \@ifx{#1\undefined}
}%
\providecommand \@ifnum [1]{%
 \ifnum #1\expandafter \@firstoftwo
 \else \expandafter \@secondoftwo
 \fi
}%
\providecommand \@ifx [1]{%
 \ifx #1\expandafter \@firstoftwo
 \else \expandafter \@secondoftwo
 \fi
}%
\providecommand \natexlab [1]{#1}%
\providecommand \enquote  [1]{``#1''}%
\providecommand \bibnamefont  [1]{#1}%
\providecommand \bibfnamefont [1]{#1}%
\providecommand \citenamefont [1]{#1}%
\providecommand \href@noop [0]{\@secondoftwo}%
\providecommand \href [0]{\begingroup \@sanitize@url \@href}%
\providecommand \@href[1]{\@@startlink{#1}\@@href}%
\providecommand \@@href[1]{\endgroup#1\@@endlink}%
\providecommand \@sanitize@url [0]{\catcode `\\12\catcode `\$12\catcode
  `\&12\catcode `\#12\catcode `\^12\catcode `\_12\catcode `\%12\relax}%
\providecommand \@@startlink[1]{}%
\providecommand \@@endlink[0]{}%
\providecommand \url  [0]{\begingroup\@sanitize@url \@url }%
\providecommand \@url [1]{\endgroup\@href {#1}{\urlprefix }}%
\providecommand \urlprefix  [0]{URL }%
\providecommand \Eprint [0]{\href }%
\providecommand \doibase [0]{https://doi.org/}%
\providecommand \selectlanguage [0]{\@gobble}%
\providecommand \bibinfo  [0]{\@secondoftwo}%
\providecommand \bibfield  [0]{\@secondoftwo}%
\providecommand \translation [1]{[#1]}%
\providecommand \BibitemOpen [0]{}%
\providecommand \bibitemStop [0]{}%
\providecommand \bibitemNoStop [0]{.\EOS\space}%
\providecommand \EOS [0]{\spacefactor3000\relax}%
\providecommand \BibitemShut  [1]{\csname bibitem#1\endcsname}%
\let\auto@bib@innerbib\@empty
\bibitem [{\citenamefont {Veres}\ \emph {et~al.}(2024)\citenamefont {Veres},
  \citenamefont {Giovannozzi},\ and\ \citenamefont {Franchetti}}]{DEV_sx_PRR}%
  \BibitemOpen
  \bibfield  {author} {\bibinfo {author} {\bibfnamefont {D.~E.}\ \bibnamefont
  {Veres}}, \bibinfo {author} {\bibfnamefont {M.}~\bibnamefont {Giovannozzi}},\
  and\ \bibinfo {author} {\bibfnamefont {G.}~\bibnamefont {Franchetti}},\
  }\bibfield  {title} {\bibinfo {title} {Exploring the potential of resonance
  islands and bent crystals for a slow extraction from circular hadron
  accelerators},\ }\href {https://doi.org/10.1103/PhysRevResearch.6.L042018}
  {\bibfield  {journal} {\bibinfo  {journal} {Phys. Rev. Res.}\ }\textbf
  {\bibinfo {volume} {6}},\ \bibinfo {pages} {L042018} (\bibinfo {year}
  {2024})}\BibitemShut {NoStop}%
\bibitem [{\citenamefont {Cappi}\ and\ \citenamefont
  {Giovannozzi}(2002)}]{PhysRevLett.88.104801}%
  \BibitemOpen
  \bibfield  {author} {\bibinfo {author} {\bibfnamefont {R.}~\bibnamefont
  {Cappi}}\ and\ \bibinfo {author} {\bibfnamefont {M.}~\bibnamefont
  {Giovannozzi}},\ }\bibfield  {title} {\bibinfo {title} {Novel method for
  multiturn extraction: Trapping charged particles in islands of phase space},\
  }\href {https://doi.org/10.1103/PhysRevLett.88.104801} {\bibfield  {journal}
  {\bibinfo  {journal} {Phys. Rev. Lett.}\ }\textbf {\bibinfo {volume} {88}},\
  \bibinfo {pages} {104801} (\bibinfo {year} {2002})}\BibitemShut {NoStop}%
\bibitem [{\citenamefont {Cappi}\ and\ \citenamefont
  {Giovannozzi}(2004)}]{PhysRevSTAB.7.024001}%
  \BibitemOpen
  \bibfield  {author} {\bibinfo {author} {\bibfnamefont {R.}~\bibnamefont
  {Cappi}}\ and\ \bibinfo {author} {\bibfnamefont {M.}~\bibnamefont
  {Giovannozzi}},\ }\bibfield  {title} {\bibinfo {title} {Multiturn extraction
  and injection by means of adiabatic capture in stable islands of phase
  space},\ }\href {https://doi.org/10.1103/PhysRevSTAB.7.024001} {\bibfield
  {journal} {\bibinfo  {journal} {Phys. Rev. ST Accel. Beams}\ }\textbf
  {\bibinfo {volume} {7}},\ \bibinfo {pages} {024001} (\bibinfo {year}
  {2004})}\BibitemShut {NoStop}%
\bibitem [{\citenamefont {Borburgh}\ \emph {et~al.}(2016)\citenamefont
  {Borburgh}, \citenamefont {Damjanovic}, \citenamefont {Gilardoni},
  \citenamefont {Giovannozzi}, \citenamefont {Hernalsteens}, \citenamefont
  {Hourican}, \citenamefont {Huschauer}, \citenamefont {Kahle}, \citenamefont
  {Le~Godec}, \citenamefont {Michels},\ and\ \citenamefont
  {Sterbini}}]{Borburgh:2137954}%
  \BibitemOpen
  \bibfield  {author} {\bibinfo {author} {\bibfnamefont {J.}~\bibnamefont
  {Borburgh}}, \bibinfo {author} {\bibfnamefont {S.}~\bibnamefont
  {Damjanovic}}, \bibinfo {author} {\bibfnamefont {S.}~\bibnamefont
  {Gilardoni}}, \bibinfo {author} {\bibfnamefont {M.}~\bibnamefont
  {Giovannozzi}}, \bibinfo {author} {\bibfnamefont {C.}~\bibnamefont
  {Hernalsteens}}, \bibinfo {author} {\bibfnamefont {M.}~\bibnamefont
  {Hourican}}, \bibinfo {author} {\bibfnamefont {A.}~\bibnamefont {Huschauer}},
  \bibinfo {author} {\bibfnamefont {K.}~\bibnamefont {Kahle}}, \bibinfo
  {author} {\bibfnamefont {G.}~\bibnamefont {Le~Godec}}, \bibinfo {author}
  {\bibfnamefont {O.}~\bibnamefont {Michels}},\ and\ \bibinfo {author}
  {\bibfnamefont {G.}~\bibnamefont {Sterbini}},\ }\bibfield  {title} {\bibinfo
  {title} {{First implementation of transversely split proton beams in the
  {CERN} Proton Synchrotron for the fixed-target physics programme}},\ }\href
  {https://doi.org/10.1209/0295-5075/113/34001} {\bibfield  {journal} {\bibinfo
   {journal} {EPL}\ }\textbf {\bibinfo {volume} {113}},\ \bibinfo {pages}
  {34001. 6 p} (\bibinfo {year} {2016})}\BibitemShut {NoStop}%
\bibitem [{\citenamefont {Huschauer}\ \emph {et~al.}(2017)\citenamefont
  {Huschauer}, \citenamefont {Blas}, \citenamefont {Borburgh}, \citenamefont
  {Damjanovic}, \citenamefont {Gilardoni}, \citenamefont {Giovannozzi},
  \citenamefont {Hourican}, \citenamefont {Kahle}, \citenamefont {Le~Godec},
  \citenamefont {Michels}, \citenamefont {Sterbini},\ and\ \citenamefont
  {Hernalsteens}}]{PhysRevAccelBeams.20.061001}%
  \BibitemOpen
  \bibfield  {author} {\bibinfo {author} {\bibfnamefont {A.}~\bibnamefont
  {Huschauer}}, \bibinfo {author} {\bibfnamefont {A.}~\bibnamefont {Blas}},
  \bibinfo {author} {\bibfnamefont {J.}~\bibnamefont {Borburgh}}, \bibinfo
  {author} {\bibfnamefont {S.}~\bibnamefont {Damjanovic}}, \bibinfo {author}
  {\bibfnamefont {S.}~\bibnamefont {Gilardoni}}, \bibinfo {author}
  {\bibfnamefont {M.}~\bibnamefont {Giovannozzi}}, \bibinfo {author}
  {\bibfnamefont {M.}~\bibnamefont {Hourican}}, \bibinfo {author}
  {\bibfnamefont {K.}~\bibnamefont {Kahle}}, \bibinfo {author} {\bibfnamefont
  {G.}~\bibnamefont {Le~Godec}}, \bibinfo {author} {\bibfnamefont
  {O.}~\bibnamefont {Michels}}, \bibinfo {author} {\bibfnamefont
  {G.}~\bibnamefont {Sterbini}},\ and\ \bibinfo {author} {\bibfnamefont
  {C.}~\bibnamefont {Hernalsteens}},\ }\bibfield  {title} {\bibinfo {title}
  {Transverse beam splitting made operational: Key features of the multiturn
  extraction at the {CERN Proton Synchrotron}},\ }\href
  {https://doi.org/10.1103/PhysRevAccelBeams.20.061001} {\bibfield  {journal}
  {\bibinfo  {journal} {Phys. Rev. Accel. Beams}\ }\textbf {\bibinfo {volume}
  {20}},\ \bibinfo {pages} {061001} (\bibinfo {year} {2017})}\BibitemShut
  {NoStop}%
\bibitem [{\citenamefont {Huschauer}\ \emph {et~al.}(2019)\citenamefont
  {Huschauer}, \citenamefont {Bartosik}, \citenamefont {Cave}, \citenamefont
  {Coly}, \citenamefont {Cotte}, \citenamefont {Damerau}, \citenamefont
  {Di~Giovanni}, \citenamefont {Gilardoni}, \citenamefont {Giovannozzi},
  \citenamefont {Kain}, \citenamefont {Koukovini-Platia}, \citenamefont
  {Mikulec}, \citenamefont {Sterbini},\ and\ \citenamefont
  {Tecker}}]{PhysRevAccelBeams.22.104002}%
  \BibitemOpen
  \bibfield  {author} {\bibinfo {author} {\bibfnamefont {A.}~\bibnamefont
  {Huschauer}}, \bibinfo {author} {\bibfnamefont {H.}~\bibnamefont {Bartosik}},
  \bibinfo {author} {\bibfnamefont {S.~C.}\ \bibnamefont {Cave}}, \bibinfo
  {author} {\bibfnamefont {M.}~\bibnamefont {Coly}}, \bibinfo {author}
  {\bibfnamefont {D.}~\bibnamefont {Cotte}}, \bibinfo {author} {\bibfnamefont
  {H.}~\bibnamefont {Damerau}}, \bibinfo {author} {\bibfnamefont {G.~P.}\
  \bibnamefont {Di~Giovanni}}, \bibinfo {author} {\bibfnamefont
  {S.}~\bibnamefont {Gilardoni}}, \bibinfo {author} {\bibfnamefont
  {M.}~\bibnamefont {Giovannozzi}}, \bibinfo {author} {\bibfnamefont
  {V.}~\bibnamefont {Kain}}, \bibinfo {author} {\bibfnamefont {E.}~\bibnamefont
  {Koukovini-Platia}}, \bibinfo {author} {\bibfnamefont {B.}~\bibnamefont
  {Mikulec}}, \bibinfo {author} {\bibfnamefont {G.}~\bibnamefont {Sterbini}},\
  and\ \bibinfo {author} {\bibfnamefont {F.}~\bibnamefont {Tecker}},\
  }\bibfield  {title} {\bibinfo {title} {{Advancing the {CERN} proton
  synchrotron multiturn extraction towards the high-intensity proton beams
  frontier}},\ }\href {https://doi.org/10.1103/PhysRevAccelBeams.22.104002}
  {\bibfield  {journal} {\bibinfo  {journal} {Phys. Rev. Accel. Beams}\
  }\textbf {\bibinfo {volume} {22}},\ \bibinfo {pages} {104002} (\bibinfo
  {year} {2019})}\BibitemShut {NoStop}%
\bibitem [{\citenamefont {Vadai}\ \emph {et~al.}(2022)\citenamefont {Vadai},
  \citenamefont {Alomainy}, \citenamefont {Damerau}, \citenamefont
  {Giovannozzi},\ and\ \citenamefont
  {Huschauer}}]{PhysRevAccelBeams.25.050101}%
  \BibitemOpen
  \bibfield  {author} {\bibinfo {author} {\bibfnamefont {M.}~\bibnamefont
  {Vadai}}, \bibinfo {author} {\bibfnamefont {A.}~\bibnamefont {Alomainy}},
  \bibinfo {author} {\bibfnamefont {H.}~\bibnamefont {Damerau}}, \bibinfo
  {author} {\bibfnamefont {M.}~\bibnamefont {Giovannozzi}},\ and\ \bibinfo
  {author} {\bibfnamefont {A.}~\bibnamefont {Huschauer}},\ }\bibfield  {title}
  {\bibinfo {title} {{Barrier bucket gymnastics and transversely split proton
  beams: Performance at the CERN Proton and Super Proton Synchrotrons}},\
  }\href {https://doi.org/10.1103/PhysRevAccelBeams.25.050101} {\bibfield
  {journal} {\bibinfo  {journal} {Phys. Rev. Accel. Beams}\ }\textbf {\bibinfo
  {volume} {25}},\ \bibinfo {pages} {050101} (\bibinfo {year}
  {2022})}\BibitemShut {NoStop}%
\bibitem [{\citenamefont {Lindhard}(1965)}]{lindhard:1965}%
  \BibitemOpen
  \bibfield  {author} {\bibinfo {author} {\bibfnamefont {J.}~\bibnamefont
  {Lindhard}},\ }\bibfield  {title} {\bibinfo {title} {Influence of crystal
  lattice on motion of energetic charged particles},\ }\href@noop {} {\bibfield
   {journal} {\bibinfo  {journal} {Mat. Fys. Medd . Dan. Vid . Selsk.}\
  }\textbf {\bibinfo {volume} {34}},\ \bibinfo {pages} {1} (\bibinfo {year}
  {1965})}\BibitemShut {NoStop}%
\bibitem [{\citenamefont {Carrigan}\ \emph {et~al.}(1990)\citenamefont
  {Carrigan}, \citenamefont {Toohig},\ and\ \citenamefont
  {Tsyganov}}]{carrigan:1990}%
  \BibitemOpen
  \bibfield  {author} {\bibinfo {author} {\bibfnamefont {R.~A.}\ \bibnamefont
  {Carrigan}}, \bibinfo {author} {\bibfnamefont {T.~E.}\ \bibnamefont
  {Toohig}},\ and\ \bibinfo {author} {\bibfnamefont {E.~N.}\ \bibnamefont
  {Tsyganov}},\ }\bibfield  {title} {\bibinfo {title} {{Beam extraction from
  TeV accelerators using channeling in bent crystals}},\ }\href
  {https://doi.org/https://doi.org/10.1016/0168-583X(90)90097-E} {\bibfield
  {journal} {\bibinfo  {journal} {Nucl. Instrum. Methods Phys. Res., Sect. B}\
  }\textbf {\bibinfo {volume} {48}},\ \bibinfo {pages} {167} (\bibinfo {year}
  {1990})}\BibitemShut {NoStop}%
\bibitem [{\citenamefont {Elsener}\ \emph {et~al.}(1996)\citenamefont
  {Elsener}, \citenamefont {Fidecaro}, \citenamefont {Gyr}, \citenamefont
  {Herr}, \citenamefont {Klem}, \citenamefont {Mikkelsen}, \citenamefont
  {M\o{}ller}, \citenamefont {Uggerh\o{}j}, \citenamefont {Vuagnin},\ and\
  \citenamefont {Weisse}}]{elsener:1996}%
  \BibitemOpen
  \bibfield  {author} {\bibinfo {author} {\bibfnamefont {K.}~\bibnamefont
  {Elsener}}, \bibinfo {author} {\bibfnamefont {G.}~\bibnamefont {Fidecaro}},
  \bibinfo {author} {\bibfnamefont {M.}~\bibnamefont {Gyr}}, \bibinfo {author}
  {\bibfnamefont {W.}~\bibnamefont {Herr}}, \bibinfo {author} {\bibfnamefont
  {J.}~\bibnamefont {Klem}}, \bibinfo {author} {\bibfnamefont {U.}~\bibnamefont
  {Mikkelsen}}, \bibinfo {author} {\bibfnamefont {S.~P.}\ \bibnamefont
  {M\o{}ller}}, \bibinfo {author} {\bibfnamefont {E.}~\bibnamefont
  {Uggerh\o{}j}}, \bibinfo {author} {\bibfnamefont {G.}~\bibnamefont
  {Vuagnin}},\ and\ \bibinfo {author} {\bibfnamefont {E.}~\bibnamefont
  {Weisse}},\ }\bibfield  {title} {\bibinfo {title} {{Proton extraction from
  the CERN SPS using bent silicon crystals}},\ }\href
  {https://doi.org/https://doi.org/10.1016/0168-583X(96)00239-X} {\bibfield
  {journal} {\bibinfo  {journal} {Nucl. Instrum. Methods Phys. Res., Sect. B}\
  }\textbf {\bibinfo {volume} {119}},\ \bibinfo {pages} {215} (\bibinfo {year}
  {1996})}\BibitemShut {NoStop}%
\bibitem [{\citenamefont {Biryukov}\ and\ \citenamefont
  {Kotov}(1997)}]{biryukov:1997}%
  \BibitemOpen
  \bibfield  {author} {\bibinfo {author} {\bibfnamefont {Y.~A.}\ \bibnamefont
  {Biryukov}, \bibfnamefont {Valery M. asnd~Chesnokov}}\ and\ \bibinfo {author}
  {\bibfnamefont {V.~I.}\ \bibnamefont {Kotov}},\ }\href@noop {} {\emph
  {\bibinfo {title} {Crystal Channeling and Its Application at High-Energy
  Accelerators}}}\ (\bibinfo  {publisher} {Springer Berlin},\ \bibinfo
  {address} {Heidelberg},\ \bibinfo {year} {1997})\ p.\ \bibinfo {pages}
  {219}\BibitemShut {NoStop}%
\bibitem [{\citenamefont {Afonin}\ \emph {et~al.}(2001)\citenamefont {Afonin},
  \citenamefont {Baranov}, \citenamefont {Biryukov}, \citenamefont {Breese},
  \citenamefont {Chepegin}, \citenamefont {Chesnokov}, \citenamefont {Guidi},
  \citenamefont {Ivanov}, \citenamefont {Kotov}, \citenamefont {Martinelli},
  \citenamefont {Scandale}, \citenamefont {Stefancich}, \citenamefont
  {Terekhov}, \citenamefont {Trbojevic}, \citenamefont {Troyanov},\ and\
  \citenamefont {Vincenzi}}]{PhysRevLett.87.094802}%
  \BibitemOpen
  \bibfield  {author} {\bibinfo {author} {\bibfnamefont {A.~G.}\ \bibnamefont
  {Afonin}}, \bibinfo {author} {\bibfnamefont {V.~T.}\ \bibnamefont {Baranov}},
  \bibinfo {author} {\bibfnamefont {V.~M.}\ \bibnamefont {Biryukov}}, \bibinfo
  {author} {\bibfnamefont {M.~B.~H.}\ \bibnamefont {Breese}}, \bibinfo {author}
  {\bibfnamefont {V.~N.}\ \bibnamefont {Chepegin}}, \bibinfo {author}
  {\bibfnamefont {Y.~A.}\ \bibnamefont {Chesnokov}}, \bibinfo {author}
  {\bibfnamefont {V.}~\bibnamefont {Guidi}}, \bibinfo {author} {\bibfnamefont
  {Y.~M.}\ \bibnamefont {Ivanov}}, \bibinfo {author} {\bibfnamefont {V.~I.}\
  \bibnamefont {Kotov}}, \bibinfo {author} {\bibfnamefont {G.}~\bibnamefont
  {Martinelli}}, \bibinfo {author} {\bibfnamefont {W.}~\bibnamefont
  {Scandale}}, \bibinfo {author} {\bibfnamefont {M.}~\bibnamefont
  {Stefancich}}, \bibinfo {author} {\bibfnamefont {V.~I.}\ \bibnamefont
  {Terekhov}}, \bibinfo {author} {\bibfnamefont {D.}~\bibnamefont {Trbojevic}},
  \bibinfo {author} {\bibfnamefont {E.~F.}\ \bibnamefont {Troyanov}},\ and\
  \bibinfo {author} {\bibfnamefont {D.}~\bibnamefont {Vincenzi}},\ }\bibfield
  {title} {\bibinfo {title} {{High-Efficiency Beam Extraction and Collimation
  Using Channeling in Very Short Bent Crystals}},\ }\href
  {https://doi.org/10.1103/PhysRevLett.87.094802} {\bibfield  {journal}
  {\bibinfo  {journal} {Phys. Rev. Lett.}\ }\textbf {\bibinfo {volume} {87}},\
  \bibinfo {pages} {094802} (\bibinfo {year} {2001})}\BibitemShut {NoStop}%
\bibitem [{\citenamefont {Bellucci}\ \emph {et~al.}(2003)\citenamefont
  {Bellucci}, \citenamefont {Biryukov}, \citenamefont {Chesnokov},
  \citenamefont {Guidi},\ and\ \citenamefont
  {Scandale}}]{PhysRevSTAB.6.033502}%
  \BibitemOpen
  \bibfield  {author} {\bibinfo {author} {\bibfnamefont {S.}~\bibnamefont
  {Bellucci}}, \bibinfo {author} {\bibfnamefont {V.~M.}\ \bibnamefont
  {Biryukov}}, \bibinfo {author} {\bibfnamefont {Y.~A.}\ \bibnamefont
  {Chesnokov}}, \bibinfo {author} {\bibfnamefont {V.}~\bibnamefont {Guidi}},\
  and\ \bibinfo {author} {\bibfnamefont {W.}~\bibnamefont {Scandale}},\
  }\bibfield  {title} {\bibinfo {title} {Making microbeams and nanobeams by
  channeling in microstructures and nanostructures},\ }\href
  {https://doi.org/10.1103/PhysRevSTAB.6.033502} {\bibfield  {journal}
  {\bibinfo  {journal} {Phys. Rev. ST Accel. Beams}\ }\textbf {\bibinfo
  {volume} {6}},\ \bibinfo {pages} {033502} (\bibinfo {year}
  {2003})}\BibitemShut {NoStop}%
\bibitem [{\citenamefont {Fliller}\ \emph {et~al.}(2005)\citenamefont
  {Fliller}, \citenamefont {Drees}, \citenamefont {Gassner}, \citenamefont
  {Hammons}, \citenamefont {McIntyre}, \citenamefont {Peggs}, \citenamefont
  {Trbojevic}, \citenamefont {Biryukov}, \citenamefont {Chesnokov},\ and\
  \citenamefont {Terekhov}}]{fliller:2005}%
  \BibitemOpen
  \bibfield  {author} {\bibinfo {author} {\bibfnamefont {R.~P.}\ \bibnamefont
  {Fliller}}, \bibinfo {author} {\bibfnamefont {A.}~\bibnamefont {Drees}},
  \bibinfo {author} {\bibfnamefont {D.}~\bibnamefont {Gassner}}, \bibinfo
  {author} {\bibfnamefont {L.}~\bibnamefont {Hammons}}, \bibinfo {author}
  {\bibfnamefont {G.}~\bibnamefont {McIntyre}}, \bibinfo {author}
  {\bibfnamefont {S.}~\bibnamefont {Peggs}}, \bibinfo {author} {\bibfnamefont
  {D.}~\bibnamefont {Trbojevic}}, \bibinfo {author} {\bibfnamefont
  {V.}~\bibnamefont {Biryukov}}, \bibinfo {author} {\bibfnamefont
  {Y.}~\bibnamefont {Chesnokov}},\ and\ \bibinfo {author} {\bibfnamefont
  {V.}~\bibnamefont {Terekhov}},\ }\bibfield  {title} {\bibinfo {title} {{RHIC
  crystal collimation}},\ }\href
  {https://doi.org/https://doi.org/10.1016/j.nimb.2005.03.004} {\bibfield
  {journal} {\bibinfo  {journal} {Nucl. Instrum. Methods Phys. Res., Sect. B}\
  }\textbf {\bibinfo {volume} {234}},\ \bibinfo {pages} {47} (\bibinfo {year}
  {2005})}\BibitemShut {NoStop}%
\bibitem [{\citenamefont {Scandale}\ \emph {et~al.}(2007)\citenamefont
  {Scandale}, \citenamefont {Still}, \citenamefont {Carnera}, \citenamefont
  {Della~Mea}, \citenamefont {De~Salvador}, \citenamefont {Milan},
  \citenamefont {Vomiero}, \citenamefont {Baricordi}, \citenamefont {Dalpiaz},
  \citenamefont {Fiorini}, \citenamefont {Guidi}, \citenamefont {Martinelli},
  \citenamefont {Mazzolari}, \citenamefont {Milan}, \citenamefont {Ambrosi},
  \citenamefont {Azzarello}, \citenamefont {Battiston}, \citenamefont
  {Bertucci}, \citenamefont {Burger}, \citenamefont {Ionica}, \citenamefont
  {Zuccon}, \citenamefont {Cavoto}, \citenamefont {Santacesaria}, \citenamefont
  {Valente}, \citenamefont {Vallazza}, \citenamefont {Afonin}, \citenamefont
  {Baranov}, \citenamefont {Chesnokov}, \citenamefont {Kotov}, \citenamefont
  {Maisheev}, \citenamefont {Yaznin}, \citenamefont {Afansiev}, \citenamefont
  {Kovalenko}, \citenamefont {Taratin}, \citenamefont {Denisov}, \citenamefont
  {Gavrikov}, \citenamefont {Ivanov}, \citenamefont {Ivochkin}, \citenamefont
  {Kosyanenko}, \citenamefont {Petrunin}, \citenamefont {Skorobogatov},
  \citenamefont {Suvorov}, \citenamefont {Bolognini}, \citenamefont {Foggetta},
  \citenamefont {Hasan},\ and\ \citenamefont {Prest}}]{PhysRevLett.98.154801}%
  \BibitemOpen
  \bibfield  {author} {\bibinfo {author} {\bibfnamefont {W.}~\bibnamefont
  {Scandale}}, \bibinfo {author} {\bibfnamefont {D.~A.}\ \bibnamefont {Still}},
  \bibinfo {author} {\bibfnamefont {A.}~\bibnamefont {Carnera}}, \bibinfo
  {author} {\bibfnamefont {G.}~\bibnamefont {Della~Mea}}, \bibinfo {author}
  {\bibfnamefont {D.}~\bibnamefont {De~Salvador}}, \bibinfo {author}
  {\bibfnamefont {R.}~\bibnamefont {Milan}}, \bibinfo {author} {\bibfnamefont
  {A.}~\bibnamefont {Vomiero}}, \bibinfo {author} {\bibfnamefont
  {S.}~\bibnamefont {Baricordi}}, \bibinfo {author} {\bibfnamefont
  {P.}~\bibnamefont {Dalpiaz}}, \bibinfo {author} {\bibfnamefont
  {M.}~\bibnamefont {Fiorini}}, \bibinfo {author} {\bibfnamefont
  {V.}~\bibnamefont {Guidi}}, \bibinfo {author} {\bibfnamefont
  {G.}~\bibnamefont {Martinelli}}, \bibinfo {author} {\bibfnamefont
  {A.}~\bibnamefont {Mazzolari}}, \bibinfo {author} {\bibfnamefont
  {E.}~\bibnamefont {Milan}}, \bibinfo {author} {\bibfnamefont
  {G.}~\bibnamefont {Ambrosi}}, \bibinfo {author} {\bibfnamefont
  {P.}~\bibnamefont {Azzarello}}, \bibinfo {author} {\bibfnamefont
  {R.}~\bibnamefont {Battiston}}, \bibinfo {author} {\bibfnamefont
  {B.}~\bibnamefont {Bertucci}}, \bibinfo {author} {\bibfnamefont {W.~J.}\
  \bibnamefont {Burger}}, \bibinfo {author} {\bibfnamefont {M.}~\bibnamefont
  {Ionica}}, \bibinfo {author} {\bibfnamefont {P.}~\bibnamefont {Zuccon}},
  \bibinfo {author} {\bibfnamefont {G.}~\bibnamefont {Cavoto}}, \bibinfo
  {author} {\bibfnamefont {R.}~\bibnamefont {Santacesaria}}, \bibinfo {author}
  {\bibfnamefont {P.}~\bibnamefont {Valente}}, \bibinfo {author} {\bibfnamefont
  {E.}~\bibnamefont {Vallazza}}, \bibinfo {author} {\bibfnamefont {A.~G.}\
  \bibnamefont {Afonin}}, \bibinfo {author} {\bibfnamefont {V.~T.}\
  \bibnamefont {Baranov}}, \bibinfo {author} {\bibfnamefont {Y.~A.}\
  \bibnamefont {Chesnokov}}, \bibinfo {author} {\bibfnamefont {V.~I.}\
  \bibnamefont {Kotov}}, \bibinfo {author} {\bibfnamefont {V.~A.}\ \bibnamefont
  {Maisheev}}, \bibinfo {author} {\bibfnamefont {I.~A.}\ \bibnamefont
  {Yaznin}}, \bibinfo {author} {\bibfnamefont {S.~V.}\ \bibnamefont
  {Afansiev}}, \bibinfo {author} {\bibfnamefont {A.~D.}\ \bibnamefont
  {Kovalenko}}, \bibinfo {author} {\bibfnamefont {A.~M.}\ \bibnamefont
  {Taratin}}, \bibinfo {author} {\bibfnamefont {A.~S.}\ \bibnamefont
  {Denisov}}, \bibinfo {author} {\bibfnamefont {Y.~A.}\ \bibnamefont
  {Gavrikov}}, \bibinfo {author} {\bibfnamefont {Y.~M.}\ \bibnamefont
  {Ivanov}}, \bibinfo {author} {\bibfnamefont {V.~G.}\ \bibnamefont
  {Ivochkin}}, \bibinfo {author} {\bibfnamefont {S.~V.}\ \bibnamefont
  {Kosyanenko}}, \bibinfo {author} {\bibfnamefont {A.~A.}\ \bibnamefont
  {Petrunin}}, \bibinfo {author} {\bibfnamefont {V.~V.}\ \bibnamefont
  {Skorobogatov}}, \bibinfo {author} {\bibfnamefont {V.~M.}\ \bibnamefont
  {Suvorov}}, \bibinfo {author} {\bibfnamefont {D.}~\bibnamefont {Bolognini}},
  \bibinfo {author} {\bibfnamefont {L.}~\bibnamefont {Foggetta}}, \bibinfo
  {author} {\bibfnamefont {S.}~\bibnamefont {Hasan}},\ and\ \bibinfo {author}
  {\bibfnamefont {M.}~\bibnamefont {Prest}},\ }\bibfield  {title} {\bibinfo
  {title} {{High-Efficiency Volume Reflection of an Ultrarelativistic Proton
  Beam with a Bent Silicon Crystal}},\ }\href
  {https://doi.org/10.1103/PhysRevLett.98.154801} {\bibfield  {journal}
  {\bibinfo  {journal} {Phys. Rev. Lett.}\ }\textbf {\bibinfo {volume} {98}},\
  \bibinfo {pages} {154801} (\bibinfo {year} {2007})}\BibitemShut {NoStop}%
\bibitem [{\citenamefont {Fliller}\ \emph {et~al.}(2006)\citenamefont
  {Fliller}, \citenamefont {Drees}, \citenamefont {Gassner}, \citenamefont
  {Hammons}, \citenamefont {McIntyre}, \citenamefont {Peggs}, \citenamefont
  {Trbojevic}, \citenamefont {Biryukov}, \citenamefont {Chesnokov},\ and\
  \citenamefont {Terekhov}}]{PhysRevSTAB.9.013501}%
  \BibitemOpen
  \bibfield  {author} {\bibinfo {author} {\bibfnamefont {R.~P.}\ \bibnamefont
  {Fliller}}, \bibinfo {author} {\bibfnamefont {A.}~\bibnamefont {Drees}},
  \bibinfo {author} {\bibfnamefont {D.}~\bibnamefont {Gassner}}, \bibinfo
  {author} {\bibfnamefont {L.}~\bibnamefont {Hammons}}, \bibinfo {author}
  {\bibfnamefont {G.}~\bibnamefont {McIntyre}}, \bibinfo {author}
  {\bibfnamefont {S.}~\bibnamefont {Peggs}}, \bibinfo {author} {\bibfnamefont
  {D.}~\bibnamefont {Trbojevic}}, \bibinfo {author} {\bibfnamefont
  {V.}~\bibnamefont {Biryukov}}, \bibinfo {author} {\bibfnamefont
  {Y.}~\bibnamefont {Chesnokov}},\ and\ \bibinfo {author} {\bibfnamefont
  {V.}~\bibnamefont {Terekhov}},\ }\bibfield  {title} {\bibinfo {title}
  {{Results of bent crystal channeling and collimation at the Relativistic
  Heavy Ion Collider}},\ }\href {https://doi.org/10.1103/PhysRevSTAB.9.013501}
  {\bibfield  {journal} {\bibinfo  {journal} {Phys. Rev. ST Accel. Beams}\
  }\textbf {\bibinfo {volume} {9}},\ \bibinfo {pages} {013501} (\bibinfo {year}
  {2006})}\BibitemShut {NoStop}%
\bibitem [{\citenamefont {Scandale}\ \emph {et~al.}(2008)\citenamefont
  {Scandale}, \citenamefont {Carnera}, \citenamefont {Della~Mea}, \citenamefont
  {De~Salvador}, \citenamefont {Milan}, \citenamefont {Vomiero}, \citenamefont
  {Baricordi}, \citenamefont {Dalpiaz}, \citenamefont {Fiorini}, \citenamefont
  {Guidi}, \citenamefont {Martinelli}, \citenamefont {Mazzolari}, \citenamefont
  {Milan}, \citenamefont {Ambrosi}, \citenamefont {Azzarello}, \citenamefont
  {Battiston}, \citenamefont {Bertucci}, \citenamefont {Burger}, \citenamefont
  {Ionica}, \citenamefont {Zuccon}, \citenamefont {Cavoto}, \citenamefont
  {Santacesaria}, \citenamefont {Valente}, \citenamefont {Vallazza},
  \citenamefont {Afonin}, \citenamefont {Baranov}, \citenamefont {Chesnokov},
  \citenamefont {Kotov}, \citenamefont {Maisheev}, \citenamefont {Yazynin},
  \citenamefont {Afanasiev}, \citenamefont {Kovalenko}, \citenamefont
  {Taratin}, \citenamefont {Denisov}, \citenamefont {Gavrikov}, \citenamefont
  {Ivanov}, \citenamefont {Ivochkin}, \citenamefont {Kosyanenko}, \citenamefont
  {Petrunin}, \citenamefont {Skorobogatov}, \citenamefont {Suvorov},
  \citenamefont {Bolognini}, \citenamefont {Foggetta}, \citenamefont {Hasan},\
  and\ \citenamefont {Prest}}]{PhysRevSTAB.11.063501}%
  \BibitemOpen
  \bibfield  {author} {\bibinfo {author} {\bibfnamefont {W.}~\bibnamefont
  {Scandale}}, \bibinfo {author} {\bibfnamefont {A.}~\bibnamefont {Carnera}},
  \bibinfo {author} {\bibfnamefont {G.}~\bibnamefont {Della~Mea}}, \bibinfo
  {author} {\bibfnamefont {D.}~\bibnamefont {De~Salvador}}, \bibinfo {author}
  {\bibfnamefont {R.}~\bibnamefont {Milan}}, \bibinfo {author} {\bibfnamefont
  {A.}~\bibnamefont {Vomiero}}, \bibinfo {author} {\bibfnamefont
  {S.}~\bibnamefont {Baricordi}}, \bibinfo {author} {\bibfnamefont
  {P.}~\bibnamefont {Dalpiaz}}, \bibinfo {author} {\bibfnamefont
  {M.}~\bibnamefont {Fiorini}}, \bibinfo {author} {\bibfnamefont
  {V.}~\bibnamefont {Guidi}}, \bibinfo {author} {\bibfnamefont
  {G.}~\bibnamefont {Martinelli}}, \bibinfo {author} {\bibfnamefont
  {A.}~\bibnamefont {Mazzolari}}, \bibinfo {author} {\bibfnamefont
  {E.}~\bibnamefont {Milan}}, \bibinfo {author} {\bibfnamefont
  {G.}~\bibnamefont {Ambrosi}}, \bibinfo {author} {\bibfnamefont
  {P.}~\bibnamefont {Azzarello}}, \bibinfo {author} {\bibfnamefont
  {R.}~\bibnamefont {Battiston}}, \bibinfo {author} {\bibfnamefont
  {B.}~\bibnamefont {Bertucci}}, \bibinfo {author} {\bibfnamefont {W.~J.}\
  \bibnamefont {Burger}}, \bibinfo {author} {\bibfnamefont {M.}~\bibnamefont
  {Ionica}}, \bibinfo {author} {\bibfnamefont {P.}~\bibnamefont {Zuccon}},
  \bibinfo {author} {\bibfnamefont {G.}~\bibnamefont {Cavoto}}, \bibinfo
  {author} {\bibfnamefont {R.}~\bibnamefont {Santacesaria}}, \bibinfo {author}
  {\bibfnamefont {P.}~\bibnamefont {Valente}}, \bibinfo {author} {\bibfnamefont
  {E.}~\bibnamefont {Vallazza}}, \bibinfo {author} {\bibfnamefont {A.~G.}\
  \bibnamefont {Afonin}}, \bibinfo {author} {\bibfnamefont {V.~T.}\
  \bibnamefont {Baranov}}, \bibinfo {author} {\bibfnamefont {Y.~A.}\
  \bibnamefont {Chesnokov}}, \bibinfo {author} {\bibfnamefont {V.~I.}\
  \bibnamefont {Kotov}}, \bibinfo {author} {\bibfnamefont {V.~A.}\ \bibnamefont
  {Maisheev}}, \bibinfo {author} {\bibfnamefont {I.~A.}\ \bibnamefont
  {Yazynin}}, \bibinfo {author} {\bibfnamefont {S.~V.}\ \bibnamefont
  {Afanasiev}}, \bibinfo {author} {\bibfnamefont {A.~D.}\ \bibnamefont
  {Kovalenko}}, \bibinfo {author} {\bibfnamefont {A.~M.}\ \bibnamefont
  {Taratin}}, \bibinfo {author} {\bibfnamefont {A.~S.}\ \bibnamefont
  {Denisov}}, \bibinfo {author} {\bibfnamefont {Y.~A.}\ \bibnamefont
  {Gavrikov}}, \bibinfo {author} {\bibfnamefont {Y.~M.}\ \bibnamefont
  {Ivanov}}, \bibinfo {author} {\bibfnamefont {V.~G.}\ \bibnamefont
  {Ivochkin}}, \bibinfo {author} {\bibfnamefont {S.~V.}\ \bibnamefont
  {Kosyanenko}}, \bibinfo {author} {\bibfnamefont {A.~A.}\ \bibnamefont
  {Petrunin}}, \bibinfo {author} {\bibfnamefont {V.~V.}\ \bibnamefont
  {Skorobogatov}}, \bibinfo {author} {\bibfnamefont {V.~M.}\ \bibnamefont
  {Suvorov}}, \bibinfo {author} {\bibfnamefont {D.}~\bibnamefont {Bolognini}},
  \bibinfo {author} {\bibfnamefont {L.}~\bibnamefont {Foggetta}}, \bibinfo
  {author} {\bibfnamefont {S.}~\bibnamefont {Hasan}},\ and\ \bibinfo {author}
  {\bibfnamefont {M.}~\bibnamefont {Prest}},\ }\bibfield  {title} {\bibinfo
  {title} {{Deflection of $400\mathrm{GeV}/c$ proton beam with bent silicon
  crystals at the CERN Super Proton Synchrotron}},\ }\href
  {https://doi.org/10.1103/PhysRevSTAB.11.063501} {\bibfield  {journal}
  {\bibinfo  {journal} {Phys. Rev. ST Accel. Beams}\ }\textbf {\bibinfo
  {volume} {11}},\ \bibinfo {pages} {063501} (\bibinfo {year}
  {2008})}\BibitemShut {NoStop}%
\bibitem [{\citenamefont {Scandale}\ \emph {et~al.}(2009)\citenamefont
  {Scandale}, \citenamefont {Vomiero}, \citenamefont {Baricordi}, \citenamefont
  {Dalpiaz}, \citenamefont {Fiorini}, \citenamefont {Guidi}, \citenamefont
  {Mazzolari}, \citenamefont {Della~Mea}, \citenamefont {Milan}, \citenamefont
  {Ambrosi}, \citenamefont {Zuccon}, \citenamefont {Bertucci}, \citenamefont
  {Burger}, \citenamefont {Duranti}, \citenamefont {Cavoto}, \citenamefont
  {Santacesaria}, \citenamefont {Valente}, \citenamefont {Luci}, \citenamefont
  {Iacoangeli}, \citenamefont {Vallazza}, \citenamefont {Afonin}, \citenamefont
  {Chesnokov}, \citenamefont {Kotov}, \citenamefont {Maisheev}, \citenamefont
  {Yazynin}, \citenamefont {Kovalenko}, \citenamefont {Taratin}, \citenamefont
  {Denisov}, \citenamefont {Gavrikov}, \citenamefont {Ivanov}, \citenamefont
  {Lapina}, \citenamefont {Malyarenko}, \citenamefont {Skorogobogatov},
  \citenamefont {Suvorov}, \citenamefont {Vavilov}, \citenamefont {Bolognini},
  \citenamefont {Hasan}, \citenamefont {Mozzanica},\ and\ \citenamefont
  {Prest}}]{PhysRevLett.102.084801}%
  \BibitemOpen
  \bibfield  {author} {\bibinfo {author} {\bibfnamefont {W.}~\bibnamefont
  {Scandale}}, \bibinfo {author} {\bibfnamefont {A.}~\bibnamefont {Vomiero}},
  \bibinfo {author} {\bibfnamefont {S.}~\bibnamefont {Baricordi}}, \bibinfo
  {author} {\bibfnamefont {P.}~\bibnamefont {Dalpiaz}}, \bibinfo {author}
  {\bibfnamefont {M.}~\bibnamefont {Fiorini}}, \bibinfo {author} {\bibfnamefont
  {V.}~\bibnamefont {Guidi}}, \bibinfo {author} {\bibfnamefont
  {A.}~\bibnamefont {Mazzolari}}, \bibinfo {author} {\bibfnamefont
  {G.}~\bibnamefont {Della~Mea}}, \bibinfo {author} {\bibfnamefont
  {R.}~\bibnamefont {Milan}}, \bibinfo {author} {\bibfnamefont
  {G.}~\bibnamefont {Ambrosi}}, \bibinfo {author} {\bibfnamefont
  {P.}~\bibnamefont {Zuccon}}, \bibinfo {author} {\bibfnamefont
  {B.}~\bibnamefont {Bertucci}}, \bibinfo {author} {\bibfnamefont
  {W.}~\bibnamefont {Burger}}, \bibinfo {author} {\bibfnamefont
  {M.}~\bibnamefont {Duranti}}, \bibinfo {author} {\bibfnamefont
  {G.}~\bibnamefont {Cavoto}}, \bibinfo {author} {\bibfnamefont
  {R.}~\bibnamefont {Santacesaria}}, \bibinfo {author} {\bibfnamefont
  {P.}~\bibnamefont {Valente}}, \bibinfo {author} {\bibfnamefont
  {C.}~\bibnamefont {Luci}}, \bibinfo {author} {\bibfnamefont {F.}~\bibnamefont
  {Iacoangeli}}, \bibinfo {author} {\bibfnamefont {E.}~\bibnamefont
  {Vallazza}}, \bibinfo {author} {\bibfnamefont {A.~G.}\ \bibnamefont
  {Afonin}}, \bibinfo {author} {\bibfnamefont {Y.~A.}\ \bibnamefont
  {Chesnokov}}, \bibinfo {author} {\bibfnamefont {V.~I.}\ \bibnamefont
  {Kotov}}, \bibinfo {author} {\bibfnamefont {V.~A.}\ \bibnamefont {Maisheev}},
  \bibinfo {author} {\bibfnamefont {I.~A.}\ \bibnamefont {Yazynin}}, \bibinfo
  {author} {\bibfnamefont {A.~D.}\ \bibnamefont {Kovalenko}}, \bibinfo {author}
  {\bibfnamefont {A.~M.}\ \bibnamefont {Taratin}}, \bibinfo {author}
  {\bibfnamefont {A.~S.}\ \bibnamefont {Denisov}}, \bibinfo {author}
  {\bibfnamefont {Y.~A.}\ \bibnamefont {Gavrikov}}, \bibinfo {author}
  {\bibfnamefont {Y.~M.}\ \bibnamefont {Ivanov}}, \bibinfo {author}
  {\bibfnamefont {L.~P.}\ \bibnamefont {Lapina}}, \bibinfo {author}
  {\bibfnamefont {L.~G.}\ \bibnamefont {Malyarenko}}, \bibinfo {author}
  {\bibfnamefont {V.~V.}\ \bibnamefont {Skorogobogatov}}, \bibinfo {author}
  {\bibfnamefont {V.~M.}\ \bibnamefont {Suvorov}}, \bibinfo {author}
  {\bibfnamefont {S.~A.}\ \bibnamefont {Vavilov}}, \bibinfo {author}
  {\bibfnamefont {D.}~\bibnamefont {Bolognini}}, \bibinfo {author}
  {\bibfnamefont {S.}~\bibnamefont {Hasan}}, \bibinfo {author} {\bibfnamefont
  {A.}~\bibnamefont {Mozzanica}},\ and\ \bibinfo {author} {\bibfnamefont
  {M.}~\bibnamefont {Prest}},\ }\bibfield  {title} {\bibinfo {title}
  {{Observation of Multiple Volume Reflection of Ultrarelativistic Protons by a
  Sequence of Several Bent Silicon Crystals}},\ }\href
  {https://doi.org/10.1103/PhysRevLett.102.084801} {\bibfield  {journal}
  {\bibinfo  {journal} {Phys. Rev. Lett.}\ }\textbf {\bibinfo {volume} {102}},\
  \bibinfo {pages} {084801} (\bibinfo {year} {2009})}\BibitemShut {NoStop}%
\bibitem [{\citenamefont {Scandale}(2010)}]{scandale:2010}%
  \BibitemOpen
  \bibfield  {author} {\bibinfo {author} {\bibfnamefont {W.}~\bibnamefont
  {Scandale}},\ }\bibfield  {title} {\bibinfo {title} {Crystal-based
  collimation in modern colliders},\ }\href
  {https://doi.org/10.1142/S0217751X1004992X} {\bibfield  {journal} {\bibinfo
  {journal} {Int. J. Mod. Phys. A}\ }\textbf {\bibinfo {volume} {25}},\
  \bibinfo {pages} {70} (\bibinfo {year} {2010})}\BibitemShut {NoStop}%
\bibitem [{\citenamefont {Scandale}\ \emph {et~al.}(2013)\citenamefont
  {Scandale}, \citenamefont {Arduini}, \citenamefont {Butcher}, \citenamefont
  {Cerutti}, \citenamefont {Gilardoni}, \citenamefont {Lari}, \citenamefont
  {Lechner}, \citenamefont {Losito}, \citenamefont {Masi}, \citenamefont
  {Mereghetti}, \citenamefont {Metral}, \citenamefont {Mirarchi}, \citenamefont
  {Montesano}, \citenamefont {Redaelli}, \citenamefont {Schoofs}, \citenamefont
  {Smirnov}, \citenamefont {Bagli}, \citenamefont {Bandiera}, \citenamefont
  {Baricordi}, \citenamefont {Dalpiaz}, \citenamefont {Guidi}, \citenamefont
  {Mazzolari}, \citenamefont {Vincenzi}, \citenamefont {Claps}, \citenamefont
  {Dabagov}, \citenamefont {Hampai}, \citenamefont {Murtas}, \citenamefont
  {Cavoto}, \citenamefont {Garattini}, \citenamefont {Iacoangeli},
  \citenamefont {Ludovici}, \citenamefont {Santacesaria}, \citenamefont
  {Valente}, \citenamefont {Galluccio}, \citenamefont {Afonin}, \citenamefont
  {Bulgakov}, \citenamefont {Chesnokov}, \citenamefont {Maisheev},
  \citenamefont {Yazynin}, \citenamefont {Kovalenko}, \citenamefont {Taratin},
  \citenamefont {Uzhinskiy}, \citenamefont {Gavrikov}, \citenamefont {Ivanov},
  \citenamefont {Lapina}, \citenamefont {Ferguson}, \citenamefont {Fulcher},
  \citenamefont {Hall}, \citenamefont {Pesaresi}, \citenamefont {Raymond},\
  and\ \citenamefont {Previtali}}]{scandale:2013}%
  \BibitemOpen
  \bibfield  {author} {\bibinfo {author} {\bibfnamefont {W.}~\bibnamefont
  {Scandale}}, \bibinfo {author} {\bibfnamefont {G.}~\bibnamefont {Arduini}},
  \bibinfo {author} {\bibfnamefont {M.}~\bibnamefont {Butcher}}, \bibinfo
  {author} {\bibfnamefont {F.}~\bibnamefont {Cerutti}}, \bibinfo {author}
  {\bibfnamefont {S.}~\bibnamefont {Gilardoni}}, \bibinfo {author}
  {\bibfnamefont {L.}~\bibnamefont {Lari}}, \bibinfo {author} {\bibfnamefont
  {A.}~\bibnamefont {Lechner}}, \bibinfo {author} {\bibfnamefont
  {R.}~\bibnamefont {Losito}}, \bibinfo {author} {\bibfnamefont
  {A.}~\bibnamefont {Masi}}, \bibinfo {author} {\bibfnamefont {A.}~\bibnamefont
  {Mereghetti}}, \bibinfo {author} {\bibfnamefont {E.}~\bibnamefont {Metral}},
  \bibinfo {author} {\bibfnamefont {D.}~\bibnamefont {Mirarchi}}, \bibinfo
  {author} {\bibfnamefont {S.}~\bibnamefont {Montesano}}, \bibinfo {author}
  {\bibfnamefont {S.}~\bibnamefont {Redaelli}}, \bibinfo {author}
  {\bibfnamefont {P.}~\bibnamefont {Schoofs}}, \bibinfo {author} {\bibfnamefont
  {G.}~\bibnamefont {Smirnov}}, \bibinfo {author} {\bibfnamefont
  {E.}~\bibnamefont {Bagli}}, \bibinfo {author} {\bibfnamefont
  {L.}~\bibnamefont {Bandiera}}, \bibinfo {author} {\bibfnamefont
  {S.}~\bibnamefont {Baricordi}}, \bibinfo {author} {\bibfnamefont
  {P.}~\bibnamefont {Dalpiaz}}, \bibinfo {author} {\bibfnamefont
  {V.}~\bibnamefont {Guidi}}, \bibinfo {author} {\bibfnamefont
  {A.}~\bibnamefont {Mazzolari}}, \bibinfo {author} {\bibfnamefont
  {D.}~\bibnamefont {Vincenzi}}, \bibinfo {author} {\bibfnamefont
  {G.}~\bibnamefont {Claps}}, \bibinfo {author} {\bibfnamefont
  {S.}~\bibnamefont {Dabagov}}, \bibinfo {author} {\bibfnamefont
  {D.}~\bibnamefont {Hampai}}, \bibinfo {author} {\bibfnamefont
  {F.}~\bibnamefont {Murtas}}, \bibinfo {author} {\bibfnamefont
  {G.}~\bibnamefont {Cavoto}}, \bibinfo {author} {\bibfnamefont
  {M.}~\bibnamefont {Garattini}}, \bibinfo {author} {\bibfnamefont
  {F.}~\bibnamefont {Iacoangeli}}, \bibinfo {author} {\bibfnamefont
  {L.}~\bibnamefont {Ludovici}}, \bibinfo {author} {\bibfnamefont
  {R.}~\bibnamefont {Santacesaria}}, \bibinfo {author} {\bibfnamefont
  {P.}~\bibnamefont {Valente}}, \bibinfo {author} {\bibfnamefont
  {F.}~\bibnamefont {Galluccio}}, \bibinfo {author} {\bibfnamefont
  {A.}~\bibnamefont {Afonin}}, \bibinfo {author} {\bibfnamefont
  {M.}~\bibnamefont {Bulgakov}}, \bibinfo {author} {\bibfnamefont
  {Y.}~\bibnamefont {Chesnokov}}, \bibinfo {author} {\bibfnamefont
  {V.}~\bibnamefont {Maisheev}}, \bibinfo {author} {\bibfnamefont
  {I.}~\bibnamefont {Yazynin}}, \bibinfo {author} {\bibfnamefont
  {A.}~\bibnamefont {Kovalenko}}, \bibinfo {author} {\bibfnamefont
  {A.}~\bibnamefont {Taratin}}, \bibinfo {author} {\bibfnamefont
  {V.}~\bibnamefont {Uzhinskiy}}, \bibinfo {author} {\bibfnamefont
  {Y.}~\bibnamefont {Gavrikov}}, \bibinfo {author} {\bibfnamefont
  {Y.}~\bibnamefont {Ivanov}}, \bibinfo {author} {\bibfnamefont
  {L.}~\bibnamefont {Lapina}}, \bibinfo {author} {\bibfnamefont
  {W.}~\bibnamefont {Ferguson}}, \bibinfo {author} {\bibfnamefont
  {J.}~\bibnamefont {Fulcher}}, \bibinfo {author} {\bibfnamefont
  {G.}~\bibnamefont {Hall}}, \bibinfo {author} {\bibfnamefont {M.}~\bibnamefont
  {Pesaresi}}, \bibinfo {author} {\bibfnamefont {M.}~\bibnamefont {Raymond}},\
  and\ \bibinfo {author} {\bibfnamefont {V.}~\bibnamefont {Previtali}},\
  }\bibfield  {title} {\bibinfo {title} {{Optimization of the crystal assisted
  collimation of the SPS beam}},\ }\href
  {https://doi.org/https://doi.org/10.1016/j.physletb.2013.08.028} {\bibfield
  {journal} {\bibinfo  {journal} {Phys. Lett. B}\ }\textbf {\bibinfo {volume}
  {726}},\ \bibinfo {pages} {182} (\bibinfo {year} {2013})}\BibitemShut
  {NoStop}%
\bibitem [{\citenamefont {Scandale}\ \emph {et~al.}(2018)\citenamefont
  {Scandale}, \citenamefont {Arduini}, \citenamefont {Cerutti}, \citenamefont
  {Garattini}, \citenamefont {Gilardoni}, \citenamefont {Masi}, \citenamefont
  {Mirarchi}, \citenamefont {Montesano}, \citenamefont {Petrucci},
  \citenamefont {Redaelli}, \citenamefont {Rossi}, \citenamefont {Breton},
  \citenamefont {Burmistrov}, \citenamefont {Dubos}, \citenamefont {Maalmi},
  \citenamefont {Natochii}, \citenamefont {Puill}, \citenamefont {Stocchi},
  \citenamefont {Sukhonos}, \citenamefont {Bagli}, \citenamefont {Bandiera},
  \citenamefont {Guidi}, \citenamefont {Mazzolari}, \citenamefont {Romagnoni},
  \citenamefont {Murtas}, \citenamefont {Addesa}, \citenamefont {Cavoto},
  \citenamefont {Iacoangeli}, \citenamefont {Galluccio}, \citenamefont
  {Afonin}, \citenamefont {Bulgakov}, \citenamefont {Chesnokov}, \citenamefont
  {Durum}, \citenamefont {Maisheev}, \citenamefont {Sandomirskiy},
  \citenamefont {Yanovich}, \citenamefont {Kolomiets}, \citenamefont
  {Kovalenko}, \citenamefont {Taratin}, \citenamefont {Smirnov}, \citenamefont
  {Denisov}, \citenamefont {Gavrikov}, \citenamefont {Ivanov}, \citenamefont
  {Lapina}, \citenamefont {Malyarenko}, \citenamefont {Skorobogatov},
  \citenamefont {Auzinger}, \citenamefont {James}, \citenamefont {Hall},
  \citenamefont {Pesaresi},\ and\ \citenamefont
  {Raymond}}]{PhysRevAccelBeams.21.014702}%
  \BibitemOpen
  \bibfield  {author} {\bibinfo {author} {\bibfnamefont {W.}~\bibnamefont
  {Scandale}}, \bibinfo {author} {\bibfnamefont {G.}~\bibnamefont {Arduini}},
  \bibinfo {author} {\bibfnamefont {F.}~\bibnamefont {Cerutti}}, \bibinfo
  {author} {\bibfnamefont {M.}~\bibnamefont {Garattini}}, \bibinfo {author}
  {\bibfnamefont {S.}~\bibnamefont {Gilardoni}}, \bibinfo {author}
  {\bibfnamefont {A.}~\bibnamefont {Masi}}, \bibinfo {author} {\bibfnamefont
  {D.}~\bibnamefont {Mirarchi}}, \bibinfo {author} {\bibfnamefont
  {S.}~\bibnamefont {Montesano}}, \bibinfo {author} {\bibfnamefont
  {S.}~\bibnamefont {Petrucci}}, \bibinfo {author} {\bibfnamefont
  {S.}~\bibnamefont {Redaelli}}, \bibinfo {author} {\bibfnamefont
  {R.}~\bibnamefont {Rossi}}, \bibinfo {author} {\bibfnamefont
  {D.}~\bibnamefont {Breton}}, \bibinfo {author} {\bibfnamefont
  {L.}~\bibnamefont {Burmistrov}}, \bibinfo {author} {\bibfnamefont
  {S.}~\bibnamefont {Dubos}}, \bibinfo {author} {\bibfnamefont
  {J.}~\bibnamefont {Maalmi}}, \bibinfo {author} {\bibfnamefont
  {A.}~\bibnamefont {Natochii}}, \bibinfo {author} {\bibfnamefont
  {V.}~\bibnamefont {Puill}}, \bibinfo {author} {\bibfnamefont
  {A.}~\bibnamefont {Stocchi}}, \bibinfo {author} {\bibfnamefont
  {D.}~\bibnamefont {Sukhonos}}, \bibinfo {author} {\bibfnamefont
  {E.}~\bibnamefont {Bagli}}, \bibinfo {author} {\bibfnamefont
  {L.}~\bibnamefont {Bandiera}}, \bibinfo {author} {\bibfnamefont
  {V.}~\bibnamefont {Guidi}}, \bibinfo {author} {\bibfnamefont
  {A.}~\bibnamefont {Mazzolari}}, \bibinfo {author} {\bibfnamefont
  {M.}~\bibnamefont {Romagnoni}}, \bibinfo {author} {\bibfnamefont
  {F.}~\bibnamefont {Murtas}}, \bibinfo {author} {\bibfnamefont
  {F.}~\bibnamefont {Addesa}}, \bibinfo {author} {\bibfnamefont
  {G.}~\bibnamefont {Cavoto}}, \bibinfo {author} {\bibfnamefont
  {F.}~\bibnamefont {Iacoangeli}}, \bibinfo {author} {\bibfnamefont
  {F.}~\bibnamefont {Galluccio}}, \bibinfo {author} {\bibfnamefont {A.~G.}\
  \bibnamefont {Afonin}}, \bibinfo {author} {\bibfnamefont {M.~K.}\
  \bibnamefont {Bulgakov}}, \bibinfo {author} {\bibfnamefont {Y.~A.}\
  \bibnamefont {Chesnokov}}, \bibinfo {author} {\bibfnamefont {A.~A.}\
  \bibnamefont {Durum}}, \bibinfo {author} {\bibfnamefont {V.~A.}\ \bibnamefont
  {Maisheev}}, \bibinfo {author} {\bibfnamefont {Y.~E.}\ \bibnamefont
  {Sandomirskiy}}, \bibinfo {author} {\bibfnamefont {A.~A.}\ \bibnamefont
  {Yanovich}}, \bibinfo {author} {\bibfnamefont {A.~A.}\ \bibnamefont
  {Kolomiets}}, \bibinfo {author} {\bibfnamefont {A.~D.}\ \bibnamefont
  {Kovalenko}}, \bibinfo {author} {\bibfnamefont {A.~M.}\ \bibnamefont
  {Taratin}}, \bibinfo {author} {\bibfnamefont {G.~I.}\ \bibnamefont
  {Smirnov}}, \bibinfo {author} {\bibfnamefont {A.~S.}\ \bibnamefont
  {Denisov}}, \bibinfo {author} {\bibfnamefont {Y.~A.}\ \bibnamefont
  {Gavrikov}}, \bibinfo {author} {\bibfnamefont {Y.~M.}\ \bibnamefont
  {Ivanov}}, \bibinfo {author} {\bibfnamefont {L.~P.}\ \bibnamefont {Lapina}},
  \bibinfo {author} {\bibfnamefont {L.~G.}\ \bibnamefont {Malyarenko}},
  \bibinfo {author} {\bibfnamefont {V.~V.}\ \bibnamefont {Skorobogatov}},
  \bibinfo {author} {\bibfnamefont {G.}~\bibnamefont {Auzinger}}, \bibinfo
  {author} {\bibfnamefont {T.}~\bibnamefont {James}}, \bibinfo {author}
  {\bibfnamefont {G.}~\bibnamefont {Hall}}, \bibinfo {author} {\bibfnamefont
  {M.}~\bibnamefont {Pesaresi}},\ and\ \bibinfo {author} {\bibfnamefont
  {M.}~\bibnamefont {Raymond}},\ }\bibfield  {title} {\bibinfo {title}
  {Comprehensive study of beam focusing by crystal devices},\ }\href
  {https://doi.org/10.1103/PhysRevAccelBeams.21.014702} {\bibfield  {journal}
  {\bibinfo  {journal} {Phys. Rev. Accel. Beams}\ }\textbf {\bibinfo {volume}
  {21}},\ \bibinfo {pages} {014702} (\bibinfo {year} {2018})}\BibitemShut
  {NoStop}%
\bibitem [{\citenamefont {Shiltsev}(2019)}]{shiltsev:2019}%
  \BibitemOpen
  \bibfield  {author} {\bibinfo {author} {\bibfnamefont {V.~D.}\ \bibnamefont
  {Shiltsev}},\ }\bibfield  {title} {\bibinfo {title} {{Experience with
  crystals at Fermilab accelerators}},\ }\href
  {https://doi.org/10.1142/S0217751X19430073} {\bibfield  {journal} {\bibinfo
  {journal} {Int. J. Mod. Phys. A}\ }\textbf {\bibinfo {volume} {34}},\
  \bibinfo {pages} {1943007} (\bibinfo {year} {2019})}\BibitemShut {NoStop}%
\bibitem [{\citenamefont {Mirarchi}\ \emph
  {et~al.}(2020{\natexlab{a}})\citenamefont {Mirarchi}, \citenamefont {Avati},
  \citenamefont {Bruce}, \citenamefont {Butcher}, \citenamefont {D'Andrea},
  \citenamefont {Di~Castro}, \citenamefont {Deile}, \citenamefont {Dziedzic},
  \citenamefont {Hiller}, \citenamefont {Jakobsen}, \citenamefont
  {Ka\ifmmode~\check{s}\else \v{s}\fi{}par}, \citenamefont {Korcyl},
  \citenamefont {Lamas}, \citenamefont {Masi}, \citenamefont {Mereghetti},
  \citenamefont {Morales}, \citenamefont {Gavrikov}, \citenamefont {Redaelli},
  \citenamefont {Ferrando}, \citenamefont {Serrano}, \citenamefont
  {Camillocci},\ and\ \citenamefont {Turini}}]{PhysRevApplied.14.064066}%
  \BibitemOpen
  \bibfield  {author} {\bibinfo {author} {\bibfnamefont {D.}~\bibnamefont
  {Mirarchi}}, \bibinfo {author} {\bibfnamefont {V.}~\bibnamefont {Avati}},
  \bibinfo {author} {\bibfnamefont {R.}~\bibnamefont {Bruce}}, \bibinfo
  {author} {\bibfnamefont {M.}~\bibnamefont {Butcher}}, \bibinfo {author}
  {\bibfnamefont {M.}~\bibnamefont {D'Andrea}}, \bibinfo {author}
  {\bibfnamefont {M.}~\bibnamefont {Di~Castro}}, \bibinfo {author}
  {\bibfnamefont {M.}~\bibnamefont {Deile}}, \bibinfo {author} {\bibfnamefont
  {B.}~\bibnamefont {Dziedzic}}, \bibinfo {author} {\bibfnamefont
  {K.}~\bibnamefont {Hiller}}, \bibinfo {author} {\bibfnamefont
  {S.}~\bibnamefont {Jakobsen}}, \bibinfo {author} {\bibfnamefont
  {J.}~\bibnamefont {Ka\ifmmode~\check{s}\else \v{s}\fi{}par}}, \bibinfo
  {author} {\bibfnamefont {K.}~\bibnamefont {Korcyl}}, \bibinfo {author}
  {\bibfnamefont {I.}~\bibnamefont {Lamas}}, \bibinfo {author} {\bibfnamefont
  {A.}~\bibnamefont {Masi}}, \bibinfo {author} {\bibfnamefont {A.}~\bibnamefont
  {Mereghetti}}, \bibinfo {author} {\bibfnamefont {H.~G.}\ \bibnamefont
  {Morales}}, \bibinfo {author} {\bibfnamefont {Y.}~\bibnamefont {Gavrikov}},
  \bibinfo {author} {\bibfnamefont {S.}~\bibnamefont {Redaelli}}, \bibinfo
  {author} {\bibfnamefont {B.~S.}\ \bibnamefont {Ferrando}}, \bibinfo {author}
  {\bibfnamefont {P.}~\bibnamefont {Serrano}}, \bibinfo {author} {\bibfnamefont
  {M.~S.}\ \bibnamefont {Camillocci}},\ and\ \bibinfo {author} {\bibfnamefont
  {N.}~\bibnamefont {Turini}},\ }\bibfield  {title} {\bibinfo {title}
  {{Reducing Beam-Related Background on Forward Physics Detectors Using Crystal
  Collimation at the Large Hadron Collider}},\ }\href
  {https://doi.org/10.1103/PhysRevApplied.14.064066} {\bibfield  {journal}
  {\bibinfo  {journal} {Phys. Rev. Appl.}\ }\textbf {\bibinfo {volume} {14}},\
  \bibinfo {pages} {064066} (\bibinfo {year} {2020}{\natexlab{a}})}\BibitemShut
  {NoStop}%
\bibitem [{\citenamefont {Scandale}\ \emph {et~al.}(2022)\citenamefont
  {Scandale}, \citenamefont {Arduini}, \citenamefont {Assmann}, \citenamefont
  {Bracco}, \citenamefont {Butcher}, \citenamefont {Cerutti}, \citenamefont
  {D'Andrea}, \citenamefont {Esposito}, \citenamefont {Garattini},
  \citenamefont {Gilardoni}, \citenamefont {Laface}, \citenamefont {Lari},
  \citenamefont {Losito}, \citenamefont {Masi}, \citenamefont {Metral},
  \citenamefont {Mirarchi}, \citenamefont {Montesano}, \citenamefont
  {Petrucci}, \citenamefont {Previtali}, \citenamefont {Redaelli},
  \citenamefont {Rossi}, \citenamefont {Schoofs}, \citenamefont {Silari},
  \citenamefont {Tlustos}, \citenamefont {Burmistrov}, \citenamefont
  {Natochii}, \citenamefont {Dubos}, \citenamefont {Puill}, \citenamefont
  {Stocchi}, \citenamefont {Bagli}, \citenamefont {Bandiera}, \citenamefont
  {Baricordi}, \citenamefont {Dalpiaz}, \citenamefont {Fiorini}, \citenamefont
  {Guidi}, \citenamefont {Mazzolari}, \citenamefont {Vincenzi}, \citenamefont
  {Addesa}, \citenamefont {Cavoto}, \citenamefont {Iacoangeli}, \citenamefont
  {Ludovici}, \citenamefont {Santacesaria}, \citenamefont {Valente},
  \citenamefont {Galluccio}, \citenamefont {Vallazza}, \citenamefont
  {Bolognini}, \citenamefont {Foggetta}, \citenamefont {Hasan}, \citenamefont
  {Lietti}, \citenamefont {Mascagna}, \citenamefont {Mattera}, \citenamefont
  {Prest}, \citenamefont {Ambrosi}, \citenamefont {Azzarello}, \citenamefont
  {Bertucci}, \citenamefont {Ionica}, \citenamefont {Battiston}, \citenamefont
  {Zuccon}, \citenamefont {Burger}, \citenamefont {Carnera}, \citenamefont
  {Della~Mea}, \citenamefont {Lombardi}, \citenamefont {De~Salvador},
  \citenamefont {Milan}, \citenamefont {Vomiero}, \citenamefont {Claps},
  \citenamefont {Dabagov}, \citenamefont {Murtas}, \citenamefont {Kovalenko},
  \citenamefont {Taratin}, \citenamefont {Uzhinskiy}, \citenamefont {Smirnov},
  \citenamefont {Denisov}, \citenamefont {Gavrikov}, \citenamefont {Ivanov},
  \citenamefont {Lapina}, \citenamefont {Malyarenko}, \citenamefont
  {Skorobogatov}, \citenamefont {Suvorov}, \citenamefont {Vavilov},
  \citenamefont {Afonin}, \citenamefont {Chesnokov}, \citenamefont {Durum},
  \citenamefont {Maisheev}, \citenamefont {Sandomirskij}, \citenamefont
  {Yanovich}, \citenamefont {Yazynin}, \citenamefont {Markiewicz},
  \citenamefont {Oriunno}, \citenamefont {Wienands}, \citenamefont {Mokhov},
  \citenamefont {Still}, \citenamefont {Auzinger}, \citenamefont {Borg},
  \citenamefont {Ferguson}, \citenamefont {Fulcher}, \citenamefont {James},
  \citenamefont {Hall}, \citenamefont {Pesaresi}, \citenamefont {Raymond},
  \citenamefont {Rose}, \citenamefont {Ryan},\ and\ \citenamefont
  {Zorba}}]{scandale:2022}%
  \BibitemOpen
  \bibfield  {author} {\bibinfo {author} {\bibfnamefont {W.}~\bibnamefont
  {Scandale}}, \bibinfo {author} {\bibfnamefont {G.}~\bibnamefont {Arduini}},
  \bibinfo {author} {\bibfnamefont {R.}~\bibnamefont {Assmann}}, \bibinfo
  {author} {\bibfnamefont {C.}~\bibnamefont {Bracco}}, \bibinfo {author}
  {\bibfnamefont {M.}~\bibnamefont {Butcher}}, \bibinfo {author} {\bibfnamefont
  {F.}~\bibnamefont {Cerutti}}, \bibinfo {author} {\bibfnamefont
  {M.}~\bibnamefont {D'Andrea}}, \bibinfo {author} {\bibfnamefont {L.~S.}\
  \bibnamefont {Esposito}}, \bibinfo {author} {\bibfnamefont {M.}~\bibnamefont
  {Garattini}}, \bibinfo {author} {\bibfnamefont {S.}~\bibnamefont
  {Gilardoni}}, \bibinfo {author} {\bibfnamefont {E.}~\bibnamefont {Laface}},
  \bibinfo {author} {\bibfnamefont {L.}~\bibnamefont {Lari}}, \bibinfo {author}
  {\bibfnamefont {R.}~\bibnamefont {Losito}}, \bibinfo {author} {\bibfnamefont
  {A.}~\bibnamefont {Masi}}, \bibinfo {author} {\bibfnamefont {E.}~\bibnamefont
  {Metral}}, \bibinfo {author} {\bibfnamefont {D.}~\bibnamefont {Mirarchi}},
  \bibinfo {author} {\bibfnamefont {S.}~\bibnamefont {Montesano}}, \bibinfo
  {author} {\bibfnamefont {S.}~\bibnamefont {Petrucci}}, \bibinfo {author}
  {\bibfnamefont {V.}~\bibnamefont {Previtali}}, \bibinfo {author}
  {\bibfnamefont {S.}~\bibnamefont {Redaelli}}, \bibinfo {author}
  {\bibfnamefont {R.}~\bibnamefont {Rossi}}, \bibinfo {author} {\bibfnamefont
  {P.}~\bibnamefont {Schoofs}}, \bibinfo {author} {\bibfnamefont
  {M.}~\bibnamefont {Silari}}, \bibinfo {author} {\bibfnamefont
  {L.}~\bibnamefont {Tlustos}}, \bibinfo {author} {\bibfnamefont
  {L.}~\bibnamefont {Burmistrov}}, \bibinfo {author} {\bibfnamefont
  {A.}~\bibnamefont {Natochii}}, \bibinfo {author} {\bibfnamefont
  {S.}~\bibnamefont {Dubos}}, \bibinfo {author} {\bibfnamefont
  {V.}~\bibnamefont {Puill}}, \bibinfo {author} {\bibfnamefont
  {A.}~\bibnamefont {Stocchi}}, \bibinfo {author} {\bibfnamefont
  {E.}~\bibnamefont {Bagli}}, \bibinfo {author} {\bibfnamefont
  {L.}~\bibnamefont {Bandiera}}, \bibinfo {author} {\bibfnamefont
  {E.}~\bibnamefont {Baricordi}}, \bibinfo {author} {\bibfnamefont
  {P.}~\bibnamefont {Dalpiaz}}, \bibinfo {author} {\bibfnamefont
  {M.}~\bibnamefont {Fiorini}}, \bibinfo {author} {\bibfnamefont
  {V.}~\bibnamefont {Guidi}}, \bibinfo {author} {\bibfnamefont
  {A.}~\bibnamefont {Mazzolari}}, \bibinfo {author} {\bibfnamefont
  {D.}~\bibnamefont {Vincenzi}}, \bibinfo {author} {\bibfnamefont
  {F.}~\bibnamefont {Addesa}}, \bibinfo {author} {\bibfnamefont
  {G.}~\bibnamefont {Cavoto}}, \bibinfo {author} {\bibfnamefont
  {F.}~\bibnamefont {Iacoangeli}}, \bibinfo {author} {\bibfnamefont
  {L.}~\bibnamefont {Ludovici}}, \bibinfo {author} {\bibfnamefont
  {R.}~\bibnamefont {Santacesaria}}, \bibinfo {author} {\bibfnamefont
  {P.}~\bibnamefont {Valente}}, \bibinfo {author} {\bibfnamefont
  {F.}~\bibnamefont {Galluccio}}, \bibinfo {author} {\bibfnamefont
  {E.}~\bibnamefont {Vallazza}}, \bibinfo {author} {\bibfnamefont
  {D.}~\bibnamefont {Bolognini}}, \bibinfo {author} {\bibfnamefont
  {L.}~\bibnamefont {Foggetta}}, \bibinfo {author} {\bibfnamefont
  {S.}~\bibnamefont {Hasan}}, \bibinfo {author} {\bibfnamefont
  {D.}~\bibnamefont {Lietti}}, \bibinfo {author} {\bibfnamefont
  {V.}~\bibnamefont {Mascagna}}, \bibinfo {author} {\bibfnamefont
  {A.}~\bibnamefont {Mattera}}, \bibinfo {author} {\bibfnamefont
  {M.}~\bibnamefont {Prest}}, \bibinfo {author} {\bibfnamefont
  {G.}~\bibnamefont {Ambrosi}}, \bibinfo {author} {\bibfnamefont
  {P.}~\bibnamefont {Azzarello}}, \bibinfo {author} {\bibfnamefont
  {B.}~\bibnamefont {Bertucci}}, \bibinfo {author} {\bibfnamefont
  {M.}~\bibnamefont {Ionica}}, \bibinfo {author} {\bibfnamefont
  {R.}~\bibnamefont {Battiston}}, \bibinfo {author} {\bibfnamefont
  {P.}~\bibnamefont {Zuccon}}, \bibinfo {author} {\bibfnamefont {W.~J.}\
  \bibnamefont {Burger}}, \bibinfo {author} {\bibfnamefont {A.}~\bibnamefont
  {Carnera}}, \bibinfo {author} {\bibfnamefont {G.}~\bibnamefont {Della~Mea}},
  \bibinfo {author} {\bibfnamefont {A.}~\bibnamefont {Lombardi}}, \bibinfo
  {author} {\bibfnamefont {D.}~\bibnamefont {De~Salvador}}, \bibinfo {author}
  {\bibfnamefont {R.}~\bibnamefont {Milan}}, \bibinfo {author} {\bibfnamefont
  {A.}~\bibnamefont {Vomiero}}, \bibinfo {author} {\bibfnamefont
  {G.}~\bibnamefont {Claps}}, \bibinfo {author} {\bibfnamefont
  {S.}~\bibnamefont {Dabagov}}, \bibinfo {author} {\bibfnamefont
  {F.}~\bibnamefont {Murtas}}, \bibinfo {author} {\bibfnamefont {A.~D.}\
  \bibnamefont {Kovalenko}}, \bibinfo {author} {\bibfnamefont {A.~M.}\
  \bibnamefont {Taratin}}, \bibinfo {author} {\bibfnamefont {V.~V.}\
  \bibnamefont {Uzhinskiy}}, \bibinfo {author} {\bibfnamefont {G.~I.}\
  \bibnamefont {Smirnov}}, \bibinfo {author} {\bibfnamefont {A.~S.}\
  \bibnamefont {Denisov}}, \bibinfo {author} {\bibfnamefont {Y.~A.}\
  \bibnamefont {Gavrikov}}, \bibinfo {author} {\bibfnamefont {Y.~M.}\
  \bibnamefont {Ivanov}}, \bibinfo {author} {\bibfnamefont {L.~P.}\
  \bibnamefont {Lapina}}, \bibinfo {author} {\bibfnamefont {L.~G.}\
  \bibnamefont {Malyarenko}}, \bibinfo {author} {\bibfnamefont {V.~V.}\
  \bibnamefont {Skorobogatov}}, \bibinfo {author} {\bibfnamefont {V.~M.}\
  \bibnamefont {Suvorov}}, \bibinfo {author} {\bibfnamefont {S.~A.}\
  \bibnamefont {Vavilov}}, \bibinfo {author} {\bibfnamefont {A.~G.}\
  \bibnamefont {Afonin}}, \bibinfo {author} {\bibfnamefont {Y.~A.}\
  \bibnamefont {Chesnokov}}, \bibinfo {author} {\bibfnamefont {A.~A.}\
  \bibnamefont {Durum}}, \bibinfo {author} {\bibfnamefont {V.~A.}\ \bibnamefont
  {Maisheev}}, \bibinfo {author} {\bibfnamefont {Y.~E.}\ \bibnamefont
  {Sandomirskij}}, \bibinfo {author} {\bibfnamefont {A.~A.}\ \bibnamefont
  {Yanovich}}, \bibinfo {author} {\bibfnamefont {I.~A.}\ \bibnamefont
  {Yazynin}}, \bibinfo {author} {\bibfnamefont {T.}~\bibnamefont {Markiewicz}},
  \bibinfo {author} {\bibfnamefont {M.}~\bibnamefont {Oriunno}}, \bibinfo
  {author} {\bibfnamefont {U.}~\bibnamefont {Wienands}}, \bibinfo {author}
  {\bibfnamefont {N.}~\bibnamefont {Mokhov}}, \bibinfo {author} {\bibfnamefont
  {D.}~\bibnamefont {Still}}, \bibinfo {author} {\bibfnamefont
  {G.}~\bibnamefont {Auzinger}}, \bibinfo {author} {\bibfnamefont
  {J.}~\bibnamefont {Borg}}, \bibinfo {author} {\bibfnamefont {W.}~\bibnamefont
  {Ferguson}}, \bibinfo {author} {\bibfnamefont {J.}~\bibnamefont {Fulcher}},
  \bibinfo {author} {\bibfnamefont {T.}~\bibnamefont {James}}, \bibinfo
  {author} {\bibfnamefont {G.}~\bibnamefont {Hall}}, \bibinfo {author}
  {\bibfnamefont {M.}~\bibnamefont {Pesaresi}}, \bibinfo {author}
  {\bibfnamefont {M.}~\bibnamefont {Raymond}}, \bibinfo {author} {\bibfnamefont
  {A.}~\bibnamefont {Rose}}, \bibinfo {author} {\bibfnamefont {M.}~\bibnamefont
  {Ryan}},\ and\ \bibinfo {author} {\bibfnamefont {O.}~\bibnamefont {Zorba}},\
  }\bibfield  {title} {\bibinfo {title} {{Feasibility of crystal-assisted
  collimation in the CERN accelerator complex}},\ }\href
  {https://doi.org/10.1142/S0217751X22300046} {\bibfield  {journal} {\bibinfo
  {journal} {Int. J. Mod. Phys. A}\ }\textbf {\bibinfo {volume} {37}},\
  \bibinfo {pages} {2230004} (\bibinfo {year} {2022})}\BibitemShut {NoStop}%
\bibitem [{\citenamefont {Scandale}\ \emph {et~al.}(2016)\citenamefont
  {Scandale}, \citenamefont {Arduini}, \citenamefont {Butcher}, \citenamefont
  {Cerutti}, \citenamefont {Garattini}, \citenamefont {Gilardoni},
  \citenamefont {Lechner}, \citenamefont {Masi}, \citenamefont {Mirarchi},
  \citenamefont {Montesano}, \citenamefont {Redaelli}, \citenamefont {Rossi},
  \citenamefont {Smirnov}, \citenamefont {Breton}, \citenamefont {Burmistrov},
  \citenamefont {Chaumat}, \citenamefont {Dubos}, \citenamefont {Maalmi},
  \citenamefont {Puill}, \citenamefont {Stocchi}, \citenamefont {Bagli},
  \citenamefont {Bandiera}, \citenamefont {Germogli}, \citenamefont {Guidi},
  \citenamefont {Mazzolari}, \citenamefont {Dabagov}, \citenamefont {Murtas},
  \citenamefont {Addesa}, \citenamefont {Cavoto}, \citenamefont {Iacoangeli},
  \citenamefont {Galluccio}, \citenamefont {Afonin}, \citenamefont {Chesnokov},
  \citenamefont {Durum}, \citenamefont {Maisheev}, \citenamefont
  {Sandomirskiy}, \citenamefont {Yanovich}, \citenamefont {Kovalenko},
  \citenamefont {Taratin}, \citenamefont {Denisov}, \citenamefont {Gavrikov},
  \citenamefont {Ivanov}, \citenamefont {Lapina}, \citenamefont {Malyarenko},
  \citenamefont {Skorobogatov}, \citenamefont {James}, \citenamefont {Hall},
  \citenamefont {Pesaresi},\ and\ \citenamefont {Raymond}}]{scandale:2016}%
  \BibitemOpen
  \bibfield  {author} {\bibinfo {author} {\bibfnamefont {W.}~\bibnamefont
  {Scandale}}, \bibinfo {author} {\bibfnamefont {G.}~\bibnamefont {Arduini}},
  \bibinfo {author} {\bibfnamefont {M.}~\bibnamefont {Butcher}}, \bibinfo
  {author} {\bibfnamefont {F.}~\bibnamefont {Cerutti}}, \bibinfo {author}
  {\bibfnamefont {M.}~\bibnamefont {Garattini}}, \bibinfo {author}
  {\bibfnamefont {S.}~\bibnamefont {Gilardoni}}, \bibinfo {author}
  {\bibfnamefont {A.}~\bibnamefont {Lechner}}, \bibinfo {author} {\bibfnamefont
  {A.}~\bibnamefont {Masi}}, \bibinfo {author} {\bibfnamefont {D.}~\bibnamefont
  {Mirarchi}}, \bibinfo {author} {\bibfnamefont {S.}~\bibnamefont {Montesano}},
  \bibinfo {author} {\bibfnamefont {S.}~\bibnamefont {Redaelli}}, \bibinfo
  {author} {\bibfnamefont {R.}~\bibnamefont {Rossi}}, \bibinfo {author}
  {\bibfnamefont {G.}~\bibnamefont {Smirnov}}, \bibinfo {author} {\bibfnamefont
  {D.}~\bibnamefont {Breton}}, \bibinfo {author} {\bibfnamefont
  {L.}~\bibnamefont {Burmistrov}}, \bibinfo {author} {\bibfnamefont
  {V.}~\bibnamefont {Chaumat}}, \bibinfo {author} {\bibfnamefont
  {S.}~\bibnamefont {Dubos}}, \bibinfo {author} {\bibfnamefont
  {J.}~\bibnamefont {Maalmi}}, \bibinfo {author} {\bibfnamefont
  {V.}~\bibnamefont {Puill}}, \bibinfo {author} {\bibfnamefont
  {A.}~\bibnamefont {Stocchi}}, \bibinfo {author} {\bibfnamefont
  {E.}~\bibnamefont {Bagli}}, \bibinfo {author} {\bibfnamefont
  {L.}~\bibnamefont {Bandiera}}, \bibinfo {author} {\bibfnamefont
  {G.}~\bibnamefont {Germogli}}, \bibinfo {author} {\bibfnamefont
  {V.}~\bibnamefont {Guidi}}, \bibinfo {author} {\bibfnamefont
  {A.}~\bibnamefont {Mazzolari}}, \bibinfo {author} {\bibfnamefont
  {S.}~\bibnamefont {Dabagov}}, \bibinfo {author} {\bibfnamefont
  {F.}~\bibnamefont {Murtas}}, \bibinfo {author} {\bibfnamefont
  {F.}~\bibnamefont {Addesa}}, \bibinfo {author} {\bibfnamefont
  {G.}~\bibnamefont {Cavoto}}, \bibinfo {author} {\bibfnamefont
  {F.}~\bibnamefont {Iacoangeli}}, \bibinfo {author} {\bibfnamefont
  {F.}~\bibnamefont {Galluccio}}, \bibinfo {author} {\bibfnamefont
  {A.}~\bibnamefont {Afonin}}, \bibinfo {author} {\bibfnamefont
  {Y.}~\bibnamefont {Chesnokov}}, \bibinfo {author} {\bibfnamefont
  {A.}~\bibnamefont {Durum}}, \bibinfo {author} {\bibfnamefont
  {V.}~\bibnamefont {Maisheev}}, \bibinfo {author} {\bibfnamefont
  {Y.}~\bibnamefont {Sandomirskiy}}, \bibinfo {author} {\bibfnamefont
  {A.}~\bibnamefont {Yanovich}}, \bibinfo {author} {\bibfnamefont
  {A.}~\bibnamefont {Kovalenko}}, \bibinfo {author} {\bibfnamefont
  {A.}~\bibnamefont {Taratin}}, \bibinfo {author} {\bibfnamefont
  {A.}~\bibnamefont {Denisov}}, \bibinfo {author} {\bibfnamefont
  {Y.}~\bibnamefont {Gavrikov}}, \bibinfo {author} {\bibfnamefont
  {Y.}~\bibnamefont {Ivanov}}, \bibinfo {author} {\bibfnamefont
  {L.}~\bibnamefont {Lapina}}, \bibinfo {author} {\bibfnamefont
  {L.}~\bibnamefont {Malyarenko}}, \bibinfo {author} {\bibfnamefont
  {V.}~\bibnamefont {Skorobogatov}}, \bibinfo {author} {\bibfnamefont
  {T.}~\bibnamefont {James}}, \bibinfo {author} {\bibfnamefont
  {G.}~\bibnamefont {Hall}}, \bibinfo {author} {\bibfnamefont {M.}~\bibnamefont
  {Pesaresi}},\ and\ \bibinfo {author} {\bibfnamefont {M.}~\bibnamefont
  {Raymond}},\ }\bibfield  {title} {\bibinfo {title} {High-efficiency
  deflection of high energy protons due to channeling along the $(110)$ axis of
  a bent silicon crystal},\ }\href
  {https://doi.org/https://doi.org/10.1016/j.physletb.2016.07.072} {\bibfield
  {journal} {\bibinfo  {journal} {Phys. Lett. B}\ }\textbf {\bibinfo {volume}
  {760}},\ \bibinfo {pages} {826} (\bibinfo {year} {2016})}\BibitemShut
  {NoStop}%
\bibitem [{\citenamefont {Redaelli}\ \emph {et~al.}(2021)\citenamefont
  {Redaelli}, \citenamefont {Butcher}, \citenamefont {Barreto}, \citenamefont
  {Losito}, \citenamefont {Masi}, \citenamefont {Mirarchi}, \citenamefont
  {Montesano}, \citenamefont {Rossi}, \citenamefont {Scandale}, \citenamefont
  {Serrano~Galvez}, \citenamefont {Valentino},\ and\ \citenamefont
  {Galluccio}}]{ion-beam-channelling}%
  \BibitemOpen
  \bibfield  {author} {\bibinfo {author} {\bibfnamefont {S.}~\bibnamefont
  {Redaelli}}, \bibinfo {author} {\bibfnamefont {M.}~\bibnamefont {Butcher}},
  \bibinfo {author} {\bibfnamefont {C.}~\bibnamefont {Barreto}}, \bibinfo
  {author} {\bibfnamefont {R.}~\bibnamefont {Losito}}, \bibinfo {author}
  {\bibfnamefont {A.}~\bibnamefont {Masi}}, \bibinfo {author} {\bibfnamefont
  {D.}~\bibnamefont {Mirarchi}}, \bibinfo {author} {\bibfnamefont
  {S.}~\bibnamefont {Montesano}}, \bibinfo {author} {\bibfnamefont
  {R.}~\bibnamefont {Rossi}}, \bibinfo {author} {\bibfnamefont
  {W.}~\bibnamefont {Scandale}}, \bibinfo {author} {\bibfnamefont
  {P.}~\bibnamefont {Serrano~Galvez}}, \bibinfo {author} {\bibfnamefont
  {G.}~\bibnamefont {Valentino}},\ and\ \bibinfo {author} {\bibfnamefont
  {F.}~\bibnamefont {Galluccio}},\ }\bibfield  {title} {\bibinfo {title} {First
  observation of ion beam channeling in bent crystals at multi-tev energies},\
  }\href {https://doi.org/10.1140/epjc/s10052-021-08927-x} {\bibfield
  {journal} {\bibinfo  {journal} {Eur. Phys. J. C}\ }\textbf {\bibinfo {volume}
  {81}},\ \bibinfo {pages} {142} (\bibinfo {year} {2021})}\BibitemShut
  {NoStop}%
\bibitem [{\citenamefont {D'Andrea}\ \emph {et~al.}(2024)\citenamefont
  {D'Andrea}, \citenamefont {Aberle}, \citenamefont {Bruce}, \citenamefont
  {Butcher}, \citenamefont {Di~Castro}, \citenamefont {Cai}, \citenamefont
  {Lamas}, \citenamefont {Masi}, \citenamefont {Mirarchi}, \citenamefont
  {Redaelli}, \citenamefont {Rossi},\ and\ \citenamefont
  {Scandale}}]{PhysRevAccelBeams.27.011002}%
  \BibitemOpen
  \bibfield  {author} {\bibinfo {author} {\bibfnamefont {M.}~\bibnamefont
  {D'Andrea}}, \bibinfo {author} {\bibfnamefont {O.}~\bibnamefont {Aberle}},
  \bibinfo {author} {\bibfnamefont {R.}~\bibnamefont {Bruce}}, \bibinfo
  {author} {\bibfnamefont {M.}~\bibnamefont {Butcher}}, \bibinfo {author}
  {\bibfnamefont {M.}~\bibnamefont {Di~Castro}}, \bibinfo {author}
  {\bibfnamefont {R.}~\bibnamefont {Cai}}, \bibinfo {author} {\bibfnamefont
  {I.}~\bibnamefont {Lamas}}, \bibinfo {author} {\bibfnamefont
  {A.}~\bibnamefont {Masi}}, \bibinfo {author} {\bibfnamefont {D.}~\bibnamefont
  {Mirarchi}}, \bibinfo {author} {\bibfnamefont {S.}~\bibnamefont {Redaelli}},
  \bibinfo {author} {\bibfnamefont {R.}~\bibnamefont {Rossi}},\ and\ \bibinfo
  {author} {\bibfnamefont {W.}~\bibnamefont {Scandale}},\ }\bibfield  {title}
  {\bibinfo {title} {{Operational performance of crystal collimation with 6.37
  $Z$ TeV Pb ion beams at the LHC}},\ }\href
  {https://doi.org/10.1103/PhysRevAccelBeams.27.011002} {\bibfield  {journal}
  {\bibinfo  {journal} {Phys. Rev. Accel. Beams}\ }\textbf {\bibinfo {volume}
  {27}},\ \bibinfo {pages} {011002} (\bibinfo {year} {2024})}\BibitemShut
  {NoStop}%
\bibitem [{\citenamefont {Redaelli}\ \emph {et~al.}(2025)\citenamefont
  {Redaelli}, \citenamefont {Aberle}, \citenamefont {Abramov}, \citenamefont
  {Bruce}, \citenamefont {Cai}, \citenamefont {Calviani}, \citenamefont
  {D'Andrea}, \citenamefont {Demassieux}, \citenamefont {Dewhurst},
  \citenamefont {Di~Castro}, \citenamefont {Esposito}, \citenamefont
  {Gilardoni}, \citenamefont {Hermes}, \citenamefont {Lindstr\"om},
  \citenamefont {Lechner}, \citenamefont {Masi}, \citenamefont {Matheson},
  \citenamefont {Mirarchi}, \citenamefont {Potoine}, \citenamefont {Ricci},
  \citenamefont {Rodin}, \citenamefont {Seidenbinder}, \citenamefont {Paiva},
  \citenamefont {Bandiera}, \citenamefont {Guidi}, \citenamefont {Mazzolari},
  \citenamefont {Romagnoni}, \citenamefont {Tamisari}, \citenamefont
  {Gavrikov},\ and\ \citenamefont {Ivanov}}]{PhysRevAccelBeams.28.051001}%
  \BibitemOpen
  \bibfield  {author} {\bibinfo {author} {\bibfnamefont {S.}~\bibnamefont
  {Redaelli}}, \bibinfo {author} {\bibfnamefont {O.}~\bibnamefont {Aberle}},
  \bibinfo {author} {\bibfnamefont {A.}~\bibnamefont {Abramov}}, \bibinfo
  {author} {\bibfnamefont {R.}~\bibnamefont {Bruce}}, \bibinfo {author}
  {\bibfnamefont {R.}~\bibnamefont {Cai}}, \bibinfo {author} {\bibfnamefont
  {M.}~\bibnamefont {Calviani}}, \bibinfo {author} {\bibfnamefont
  {M.}~\bibnamefont {D'Andrea}}, \bibinfo {author} {\bibfnamefont
  {Q.}~\bibnamefont {Demassieux}}, \bibinfo {author} {\bibfnamefont
  {K.}~\bibnamefont {Dewhurst}}, \bibinfo {author} {\bibfnamefont
  {M.}~\bibnamefont {Di~Castro}}, \bibinfo {author} {\bibfnamefont {L.~S.}\
  \bibnamefont {Esposito}}, \bibinfo {author} {\bibfnamefont {S.}~\bibnamefont
  {Gilardoni}}, \bibinfo {author} {\bibfnamefont {P.~D.}\ \bibnamefont
  {Hermes}}, \bibinfo {author} {\bibfnamefont {B.}~\bibnamefont {Lindstr\"om}},
  \bibinfo {author} {\bibfnamefont {A.}~\bibnamefont {Lechner}}, \bibinfo
  {author} {\bibfnamefont {A.}~\bibnamefont {Masi}}, \bibinfo {author}
  {\bibfnamefont {E.}~\bibnamefont {Matheson}}, \bibinfo {author}
  {\bibfnamefont {D.}~\bibnamefont {Mirarchi}}, \bibinfo {author}
  {\bibfnamefont {J.-B.}\ \bibnamefont {Potoine}}, \bibinfo {author}
  {\bibfnamefont {G.}~\bibnamefont {Ricci}}, \bibinfo {author} {\bibfnamefont
  {V.}~\bibnamefont {Rodin}}, \bibinfo {author} {\bibfnamefont
  {R.}~\bibnamefont {Seidenbinder}}, \bibinfo {author} {\bibfnamefont {S.~S.}\
  \bibnamefont {Paiva}}, \bibinfo {author} {\bibfnamefont {L.}~\bibnamefont
  {Bandiera}}, \bibinfo {author} {\bibfnamefont {V.}~\bibnamefont {Guidi}},
  \bibinfo {author} {\bibfnamefont {A.}~\bibnamefont {Mazzolari}}, \bibinfo
  {author} {\bibfnamefont {M.}~\bibnamefont {Romagnoni}}, \bibinfo {author}
  {\bibfnamefont {M.}~\bibnamefont {Tamisari}}, \bibinfo {author}
  {\bibfnamefont {Y.}~\bibnamefont {Gavrikov}},\ and\ \bibinfo {author}
  {\bibfnamefont {Y.}~\bibnamefont {Ivanov}},\ }\bibfield  {title} {\bibinfo
  {title} {Crystal collimation of heavy-ion beams at the large hadron
  collider},\ }\href {https://doi.org/10.1103/PhysRevAccelBeams.28.051001}
  {\bibfield  {journal} {\bibinfo  {journal} {Phys. Rev. Accel. Beams}\
  }\textbf {\bibinfo {volume} {28}},\ \bibinfo {pages} {051001} (\bibinfo
  {year} {2025})}\BibitemShut {NoStop}%
\bibitem [{\citenamefont {Fomin}\ \emph {et~al.}(2017)\citenamefont {Fomin},
  \citenamefont {Korchin}, \citenamefont {Stocchi}, \citenamefont {Bezshyyko},
  \citenamefont {Burmistrov}, \citenamefont {Fomin}, \citenamefont {Kirillin},
  \citenamefont {Massacrier}, \citenamefont {Natochii}, \citenamefont {Robbe},
  \citenamefont {Scandale},\ and\ \citenamefont {Shul'ga}}]{fomin:2017}%
  \BibitemOpen
  \bibfield  {author} {\bibinfo {author} {\bibfnamefont {A.~S.}\ \bibnamefont
  {Fomin}}, \bibinfo {author} {\bibfnamefont {A.~Y.}\ \bibnamefont {Korchin}},
  \bibinfo {author} {\bibfnamefont {A.}~\bibnamefont {Stocchi}}, \bibinfo
  {author} {\bibfnamefont {O.~A.}\ \bibnamefont {Bezshyyko}}, \bibinfo {author}
  {\bibfnamefont {L.}~\bibnamefont {Burmistrov}}, \bibinfo {author}
  {\bibfnamefont {S.~P.}\ \bibnamefont {Fomin}}, \bibinfo {author}
  {\bibfnamefont {I.~V.}\ \bibnamefont {Kirillin}}, \bibinfo {author}
  {\bibfnamefont {L.}~\bibnamefont {Massacrier}}, \bibinfo {author}
  {\bibfnamefont {A.}~\bibnamefont {Natochii}}, \bibinfo {author}
  {\bibfnamefont {P.}~\bibnamefont {Robbe}}, \bibinfo {author} {\bibfnamefont
  {W.}~\bibnamefont {Scandale}},\ and\ \bibinfo {author} {\bibfnamefont
  {N.~F.}\ \bibnamefont {Shul'ga}},\ }\bibfield  {title} {\bibinfo {title}
  {{Feasibility of measuring the magnetic dipole moments of the charm baryons
  at the LHC using bent crystals}},\ }\href
  {https://doi.org/10.1007/JHEP08(2017)120} {\bibfield  {journal} {\bibinfo
  {journal} {J. High Energy Phys.}\ }\textbf {\bibinfo {volume} {2017}}\bibinfo
   {number} { (8)},\ \bibinfo {pages} {120}}\BibitemShut {NoStop}%
\bibitem [{\citenamefont {Bagli}\ \emph {et~al.}(2017)\citenamefont {Bagli},
  \citenamefont {Bandiera}, \citenamefont {Cavoto}, \citenamefont {Guidi},
  \citenamefont {Henry}, \citenamefont {Marangotto}, \citenamefont
  {Martinez~Vidal}, \citenamefont {Mazzolari}, \citenamefont {Merli},
  \citenamefont {Neri},\ and\ \citenamefont {Ruiz~Vidal}}]{bagli:2017}%
  \BibitemOpen
\bibfield  {number} {  }\bibfield  {author} {\bibinfo {author} {\bibfnamefont
  {E.}~\bibnamefont {Bagli}}, \bibinfo {author} {\bibfnamefont
  {L.}~\bibnamefont {Bandiera}}, \bibinfo {author} {\bibfnamefont
  {G.}~\bibnamefont {Cavoto}}, \bibinfo {author} {\bibfnamefont
  {V.}~\bibnamefont {Guidi}}, \bibinfo {author} {\bibfnamefont
  {L.}~\bibnamefont {Henry}}, \bibinfo {author} {\bibfnamefont
  {D.}~\bibnamefont {Marangotto}}, \bibinfo {author} {\bibfnamefont
  {F.}~\bibnamefont {Martinez~Vidal}}, \bibinfo {author} {\bibfnamefont
  {A.}~\bibnamefont {Mazzolari}}, \bibinfo {author} {\bibfnamefont
  {A.}~\bibnamefont {Merli}}, \bibinfo {author} {\bibfnamefont
  {N.}~\bibnamefont {Neri}},\ and\ \bibinfo {author} {\bibfnamefont
  {J.}~\bibnamefont {Ruiz~Vidal}},\ }\bibfield  {title} {\bibinfo {title}
  {{Electromagnetic dipole moments of charged baryons with bent crystals at the
  LHC}},\ }\href {https://doi.org/10.1140/epjc/s10052-017-5400-x} {\bibfield
  {journal} {\bibinfo  {journal} {Eur. Phys. J. C}\ }\textbf {\bibinfo {volume}
  {77}},\ \bibinfo {pages} {828} (\bibinfo {year} {2017})}\BibitemShut
  {NoStop}%
\bibitem [{\citenamefont {Botella}\ \emph {et~al.}(2017)\citenamefont
  {Botella}, \citenamefont {Garcia~Martin}, \citenamefont {Marangotto},
  \citenamefont {Martinez~Vidal}, \citenamefont {Merli}, \citenamefont {Neri},
  \citenamefont {Oyanguren},\ and\ \citenamefont {Ruiz~Vidal}}]{botella:2017}%
  \BibitemOpen
  \bibfield  {author} {\bibinfo {author} {\bibfnamefont {F.~J.}\ \bibnamefont
  {Botella}}, \bibinfo {author} {\bibfnamefont {L.~M.}\ \bibnamefont
  {Garcia~Martin}}, \bibinfo {author} {\bibfnamefont {D.}~\bibnamefont
  {Marangotto}}, \bibinfo {author} {\bibfnamefont {F.}~\bibnamefont
  {Martinez~Vidal}}, \bibinfo {author} {\bibfnamefont {A.}~\bibnamefont
  {Merli}}, \bibinfo {author} {\bibfnamefont {N.}~\bibnamefont {Neri}},
  \bibinfo {author} {\bibfnamefont {A.}~\bibnamefont {Oyanguren}},\ and\
  \bibinfo {author} {\bibfnamefont {J.}~\bibnamefont {Ruiz~Vidal}},\ }\bibfield
   {title} {\bibinfo {title} {{On the search for the electric dipole moment of
  strange and charm baryons at LHC}},\ }\href
  {https://doi.org/10.1140/epjc/s10052-017-4679-y} {\bibfield  {journal}
  {\bibinfo  {journal} {Eur. Phys. J. C}\ }\textbf {\bibinfo {volume} {77}},\
  \bibinfo {pages} {181} (\bibinfo {year} {2017})}\BibitemShut {NoStop}%
\bibitem [{\citenamefont {Redaelli}\ \emph {et~al.}(2018)\citenamefont
  {Redaelli}, \citenamefont {Ferro-Luzzi},\ and\ \citenamefont
  {Hadjidakis}}]{redaelli:ipac18-tupaf045}%
  \BibitemOpen
  \bibfield  {author} {\bibinfo {author} {\bibfnamefont {S.}~\bibnamefont
  {Redaelli}}, \bibinfo {author} {\bibfnamefont {M.}~\bibnamefont
  {Ferro-Luzzi}},\ and\ \bibinfo {author} {\bibfnamefont {C.}~\bibnamefont
  {Hadjidakis}},\ }\bibfield  {title} {{\selectlanguage {english}\bibinfo
  {title} {{Studies for Future Fixed-Target Experiments at the LHC in the
  Framework of the CERN Physics Beyond Colliders Study}}},\ }in\ \href
  {https://doi.org/10.18429/JACoW-IPAC2018-TUPAF045} {{\selectlanguage
  {english}\emph {\bibinfo {booktitle} {Proc. IPAC'18}}}}\ (\bibinfo
  {publisher} {JACoW Publishing, Geneva, Switzerland},\ \bibinfo {year}
  {2018})\ pp.\ \bibinfo {pages} {798--801}\BibitemShut {NoStop}%
\bibitem [{\citenamefont {Fu}\ \emph {et~al.}(2019)\citenamefont {Fu},
  \citenamefont {Giorgi}, \citenamefont {Henry}, \citenamefont {Marangotto},
  \citenamefont {Vidal}, \citenamefont {Merli}, \citenamefont {Neri},\ and\
  \citenamefont {Vidal}}]{PhysRevLett.123.011801}%
  \BibitemOpen
  \bibfield  {author} {\bibinfo {author} {\bibfnamefont {J.}~\bibnamefont
  {Fu}}, \bibinfo {author} {\bibfnamefont {M.~A.}\ \bibnamefont {Giorgi}},
  \bibinfo {author} {\bibfnamefont {L.}~\bibnamefont {Henry}}, \bibinfo
  {author} {\bibfnamefont {D.}~\bibnamefont {Marangotto}}, \bibinfo {author}
  {\bibfnamefont {F.~M.}\ \bibnamefont {Vidal}}, \bibinfo {author}
  {\bibfnamefont {A.}~\bibnamefont {Merli}}, \bibinfo {author} {\bibfnamefont
  {N.}~\bibnamefont {Neri}},\ and\ \bibinfo {author} {\bibfnamefont {J.~R.}\
  \bibnamefont {Vidal}},\ }\bibfield  {title} {\bibinfo {title} {{Novel Method
  for the Direct Measurement of the $\tau$ Lepton Dipole Moments}},\ }\href
  {https://doi.org/10.1103/PhysRevLett.123.011801} {\bibfield  {journal}
  {\bibinfo  {journal} {Phys. Rev. Lett.}\ }\textbf {\bibinfo {volume} {123}},\
  \bibinfo {pages} {011801} (\bibinfo {year} {2019})}\BibitemShut {NoStop}%
\bibitem [{\citenamefont {Fomin}\ \emph {et~al.}(2019)\citenamefont {Fomin},
  \citenamefont {Korchin}, \citenamefont {Stocchi}, \citenamefont {Barsuk},\
  and\ \citenamefont {Robbe}}]{fomin:2019}%
  \BibitemOpen
  \bibfield  {author} {\bibinfo {author} {\bibfnamefont {A.~S.}\ \bibnamefont
  {Fomin}}, \bibinfo {author} {\bibfnamefont {A.~Y.}\ \bibnamefont {Korchin}},
  \bibinfo {author} {\bibfnamefont {A.}~\bibnamefont {Stocchi}}, \bibinfo
  {author} {\bibfnamefont {S.}~\bibnamefont {Barsuk}},\ and\ \bibinfo {author}
  {\bibfnamefont {P.}~\bibnamefont {Robbe}},\ }\bibfield  {title} {\bibinfo
  {title} {{Feasibility of $\tau$-lepton electromagnetic dipole moments
  measurement using bent crystal at the LHC}},\ }\href
  {https://doi.org/10.1007/JHEP03(2019)156} {\bibfield  {journal} {\bibinfo
  {journal} {J. High Energy Phys.}\ }\textbf {\bibinfo {volume} {2019}}\bibinfo
   {number} { (3)},\ \bibinfo {pages} {156}}\BibitemShut {NoStop}%
\bibitem [{\citenamefont {Mirarchi}\ \emph
  {et~al.}(2020{\natexlab{b}})\citenamefont {Mirarchi}, \citenamefont {Fomin},
  \citenamefont {Redaelli},\ and\ \citenamefont {Scandale}}]{mirarchi:2020}%
  \BibitemOpen
\bibfield  {number} {  }\bibfield  {author} {\bibinfo {author} {\bibfnamefont
  {D.}~\bibnamefont {Mirarchi}}, \bibinfo {author} {\bibfnamefont {A.~S.}\
  \bibnamefont {Fomin}}, \bibinfo {author} {\bibfnamefont {S.}~\bibnamefont
  {Redaelli}},\ and\ \bibinfo {author} {\bibfnamefont {W.}~\bibnamefont
  {Scandale}},\ }\bibfield  {title} {\bibinfo {title} {{Layouts for
  fixed-target experiments and dipole moment measurements of short-lived
  baryons using bent crystals at the LHC}},\ }\href
  {https://doi.org/10.1140/epjc/s10052-020-08466-x} {\bibfield  {journal}
  {\bibinfo  {journal} {Eur. Phys. J. C}\ }\textbf {\bibinfo {volume} {80}},\
  \bibinfo {pages} {929} (\bibinfo {year} {2020}{\natexlab{b}})}\BibitemShut
  {NoStop}%
\bibitem [{\citenamefont {Dewhurst}\ \emph {et~al.}(2023)\citenamefont
  {Dewhurst}, \citenamefont {Mirarchi}, \citenamefont {Patecki}, \citenamefont
  {D'Andrea}, \citenamefont {Hermes},\ and\ \citenamefont
  {Redaelli}}]{Dewhurst:2023cth}%
  \BibitemOpen
  \bibfield  {author} {\bibinfo {author} {\bibfnamefont {K.~A.}\ \bibnamefont
  {Dewhurst}}, \bibinfo {author} {\bibfnamefont {D.}~\bibnamefont {Mirarchi}},
  \bibinfo {author} {\bibfnamefont {M.}~\bibnamefont {Patecki}}, \bibinfo
  {author} {\bibfnamefont {M.}~\bibnamefont {D'Andrea}}, \bibinfo {author}
  {\bibfnamefont {P.}~\bibnamefont {Hermes}},\ and\ \bibinfo {author}
  {\bibfnamefont {S.}~\bibnamefont {Redaelli}},\ }\bibfield  {title} {\bibinfo
  {title} {{Performance of a double-crystal setup for LHC fixed-target
  experiments}},\ }\href {https://doi.org/10.18429/JACoW-IPAC2023-MOPL048}
  {\bibfield  {journal} {\bibinfo  {journal} {JACoW}\ }\textbf {\bibinfo
  {volume} {IPAC2023}},\ \bibinfo {pages} {MOPL048} (\bibinfo {year}
  {2023})}\BibitemShut {NoStop}%
\bibitem [{\citenamefont {Barschel}\ \emph {et~al.}(2019)\citenamefont
  {Barschel}, \citenamefont {Bernhard}, \citenamefont {Bersani}, \citenamefont
  {Boscolo~Meneguolo}, \citenamefont {Bruce}, \citenamefont {Calviani},
  \citenamefont {Carassiti}, \citenamefont {Cerutti}, \citenamefont
  {Chiggiato}, \citenamefont {Ciullo}, \citenamefont {Di~Nezza}, \citenamefont
  {Ferro-Luzzi}, \citenamefont {Fomin}, \citenamefont {Galluccio},
  \citenamefont {Garattini}, \citenamefont {Giovannozzi}, \citenamefont
  {Hadjidakis}, \citenamefont {Kurepin}, \citenamefont {Kurepin}, \citenamefont
  {Lenisa}, \citenamefont {Macrì}, \citenamefont {Martinez~Vidal},
  \citenamefont {Massacrier}, \citenamefont {Mazzolari}, \citenamefont
  {Mereghetti}, \citenamefont {Merli}, \citenamefont {Mether}, \citenamefont
  {Mirarchi}, \citenamefont {Neri}, \citenamefont {Orth}, \citenamefont
  {Pappalardo}, \citenamefont {Poland}, \citenamefont {Popovic}, \citenamefont
  {Pressard}, \citenamefont {Redaelli}, \citenamefont {Robbe}, \citenamefont
  {Rossi}, \citenamefont {Rumolo}, \citenamefont {Salvant}, \citenamefont
  {Scandale}, \citenamefont {Steffens}, \citenamefont {Stocchi}, \citenamefont
  {Topilskaya},\ and\ \citenamefont {Vollinger}}]{Barschel:2653780}%
  \BibitemOpen
  \bibfield  {author} {\bibinfo {author} {\bibfnamefont {C.}~\bibnamefont
  {Barschel}}, \bibinfo {author} {\bibfnamefont {J.}~\bibnamefont {Bernhard}},
  \bibinfo {author} {\bibfnamefont {A.}~\bibnamefont {Bersani}}, \bibinfo
  {author} {\bibfnamefont {C.}~\bibnamefont {Boscolo~Meneguolo}}, \bibinfo
  {author} {\bibfnamefont {R.}~\bibnamefont {Bruce}}, \bibinfo {author}
  {\bibfnamefont {M.}~\bibnamefont {Calviani}}, \bibinfo {author}
  {\bibfnamefont {V.}~\bibnamefont {Carassiti}}, \bibinfo {author}
  {\bibfnamefont {F.}~\bibnamefont {Cerutti}}, \bibinfo {author} {\bibfnamefont
  {P.}~\bibnamefont {Chiggiato}}, \bibinfo {author} {\bibfnamefont
  {G.}~\bibnamefont {Ciullo}}, \bibinfo {author} {\bibfnamefont
  {P.}~\bibnamefont {Di~Nezza}}, \bibinfo {author} {\bibfnamefont
  {M.}~\bibnamefont {Ferro-Luzzi}}, \bibinfo {author} {\bibfnamefont
  {A.}~\bibnamefont {Fomin}}, \bibinfo {author} {\bibfnamefont
  {F.}~\bibnamefont {Galluccio}}, \bibinfo {author} {\bibfnamefont
  {M.}~\bibnamefont {Garattini}}, \bibinfo {author} {\bibfnamefont
  {M.}~\bibnamefont {Giovannozzi}}, \bibinfo {author} {\bibfnamefont
  {C.}~\bibnamefont {Hadjidakis}}, \bibinfo {author} {\bibfnamefont
  {A.}~\bibnamefont {Kurepin}}, \bibinfo {author} {\bibfnamefont
  {N.}~\bibnamefont {Kurepin}}, \bibinfo {author} {\bibfnamefont
  {P.}~\bibnamefont {Lenisa}}, \bibinfo {author} {\bibfnamefont
  {M.}~\bibnamefont {Macrì}}, \bibinfo {author} {\bibfnamefont
  {F.}~\bibnamefont {Martinez~Vidal}}, \bibinfo {author} {\bibfnamefont
  {L.~M.}\ \bibnamefont {Massacrier}}, \bibinfo {author} {\bibfnamefont
  {A.}~\bibnamefont {Mazzolari}}, \bibinfo {author} {\bibfnamefont
  {A.}~\bibnamefont {Mereghetti}}, \bibinfo {author} {\bibfnamefont
  {A.}~\bibnamefont {Merli}}, \bibinfo {author} {\bibfnamefont
  {L.}~\bibnamefont {Mether}}, \bibinfo {author} {\bibfnamefont
  {D.}~\bibnamefont {Mirarchi}}, \bibinfo {author} {\bibfnamefont
  {N.}~\bibnamefont {Neri}}, \bibinfo {author} {\bibfnamefont {H.}~\bibnamefont
  {Orth}}, \bibinfo {author} {\bibfnamefont {L.~L.}\ \bibnamefont
  {Pappalardo}}, \bibinfo {author} {\bibfnamefont {K.~L.}\ \bibnamefont
  {Poland}}, \bibinfo {author} {\bibfnamefont {B.~K.}\ \bibnamefont {Popovic}},
  \bibinfo {author} {\bibfnamefont {K.}~\bibnamefont {Pressard}}, \bibinfo
  {author} {\bibfnamefont {S.}~\bibnamefont {Redaelli}}, \bibinfo {author}
  {\bibfnamefont {P.}~\bibnamefont {Robbe}}, \bibinfo {author} {\bibfnamefont
  {R.}~\bibnamefont {Rossi}}, \bibinfo {author} {\bibfnamefont
  {G.}~\bibnamefont {Rumolo}}, \bibinfo {author} {\bibfnamefont
  {B.}~\bibnamefont {Salvant}}, \bibinfo {author} {\bibfnamefont
  {W.}~\bibnamefont {Scandale}}, \bibinfo {author} {\bibfnamefont
  {E.}~\bibnamefont {Steffens}}, \bibinfo {author} {\bibfnamefont
  {A.}~\bibnamefont {Stocchi}}, \bibinfo {author} {\bibfnamefont
  {N.}~\bibnamefont {Topilskaya}},\ and\ \bibinfo {author} {\bibfnamefont
  {C.}~\bibnamefont {Vollinger}},\ }\href
  {https://doi.org/10.23731/CYRM-2020-004} {\emph {\bibinfo {title} {{LHC fixed
  target experiments}}}},\ \bibinfo {type} {Tech. Rep.}\ (\bibinfo
  {institution} {CERN},\ \bibinfo {address} {Geneva},\ \bibinfo {year}
  {2019})\BibitemShut {NoStop}%
\bibitem [{\citenamefont {Jaeckel}\ \emph {et~al.}(2020)\citenamefont
  {Jaeckel}, \citenamefont {Lamont},\ and\ \citenamefont
  {Vall{\'e}e}}]{jaeckel:2020}%
  \BibitemOpen
  \bibfield  {author} {\bibinfo {author} {\bibfnamefont {J.}~\bibnamefont
  {Jaeckel}}, \bibinfo {author} {\bibfnamefont {M.}~\bibnamefont {Lamont}},\
  and\ \bibinfo {author} {\bibfnamefont {C.}~\bibnamefont {Vall{\'e}e}},\
  }\bibfield  {title} {\bibinfo {title} {The quest for new physics with the
  physics beyond colliders programme},\ }\href
  {https://doi.org/10.1038/s41567-020-0838-4} {\bibfield  {journal} {\bibinfo
  {journal} {Nat. Phys.}\ }\textbf {\bibinfo {volume} {16}},\ \bibinfo {pages}
  {393} (\bibinfo {year} {2020})}\BibitemShut {NoStop}%
\bibitem [{pbc(2023)}]{pbc}%
  \BibitemOpen
  \href@noop {} {\bibinfo {title} {{The Physics Beyond Colliders Study
  Group}}},\ \bibinfo {howpublished} {Available at
  \href{https://pbc.web.cern.ch/}{https://pbc.web.cern.ch/}} (\bibinfo {year}
  {2023})\BibitemShut {NoStop}%
\bibitem [{\citenamefont {Veres}\ \emph
  {et~al.}(2025{\natexlab{a}})\citenamefont {Veres}, \citenamefont
  {Giovannozzi},\ and\ \citenamefont {Franchetti}}]{DEV_sx_NIMA}%
  \BibitemOpen
  \bibfield  {author} {\bibinfo {author} {\bibfnamefont {D.~E.}\ \bibnamefont
  {Veres}}, \bibinfo {author} {\bibfnamefont {M.}~\bibnamefont {Giovannozzi}},\
  and\ \bibinfo {author} {\bibfnamefont {G.}~\bibnamefont {Franchetti}},\
  }\bibfield  {title} {\bibinfo {title} {An innovative method for slow
  extraction in circular hadron accelerators with resonance islands and bent
  crystals},\ }\href
  {https://doi.org/https://doi.org/10.1016/j.nima.2025.170286} {\bibfield
  {journal} {\bibinfo  {journal} {Nucl. Instrum. Methods Phys. Res., Sect. A}\
  ,\ \bibinfo {pages} {170286}} (\bibinfo {year}
  {2025}{\natexlab{a}})}\BibitemShut {NoStop}%
\bibitem [{\citenamefont {Tuck}\ and\ \citenamefont {Teng}(1951)}]{Tuck:1951}%
  \BibitemOpen
  \bibfield  {author} {\bibinfo {author} {\bibfnamefont {J.~L.}\ \bibnamefont
  {Tuck}}\ and\ \bibinfo {author} {\bibfnamefont {L.~C.}\ \bibnamefont
  {Teng}},\ }\bibfield  {title} {\bibinfo {title} {{Regenerative Deflector for
  Synchrocyclotron}},\ }\href@noop {} {\bibfield  {journal} {\bibinfo
  {journal} {Phys. Rev.}\ }\textbf {\bibinfo {volume} {81}},\ \bibinfo {pages}
  {305} (\bibinfo {year} {1951})}\BibitemShut {NoStop}%
\bibitem [{\citenamefont {Le~Couteur}(1951)}]{LeCouteur:1951}%
  \BibitemOpen
  \bibfield  {author} {\bibinfo {author} {\bibfnamefont {K.}~\bibnamefont
  {Le~Couteur}},\ }\bibfield  {title} {\bibinfo {title} {The regenerative
  deflector for synchro-cyclotrons},\ }\href@noop {} {\bibfield  {journal}
  {\bibinfo  {journal} {Proc. Phys. Soc. London, Sect. B}\ }\textbf {\bibinfo
  {volume} {64}},\ \bibinfo {pages} {1073} (\bibinfo {year}
  {1951})}\BibitemShut {NoStop}%
\bibitem [{\citenamefont {Gordon}\ and\ \citenamefont
  {Welton}(1958)}]{gordon:1958}%
  \BibitemOpen
  \bibfield  {author} {\bibinfo {author} {\bibfnamefont {M.~M.}\ \bibnamefont
  {Gordon}}\ and\ \bibinfo {author} {\bibfnamefont {T.~A.}\ \bibnamefont
  {Welton}},\ }\bibfield  {title} {\bibinfo {title} {The 8/4 resonance and beam
  extraction from the avf cyclotron},\ }\href@noop {} {\bibfield  {journal}
  {\bibinfo  {journal} {Bull. Am. Phys. Soc. 3 (1958) 57}\ }\textbf {\bibinfo
  {volume} {3}},\ \bibinfo {pages} {57} (\bibinfo {year} {1958})}\BibitemShut
  {NoStop}%
\bibitem [{\citenamefont {Hammer}\ and\ \citenamefont
  {Laslett}(2004)}]{hammer:1961}%
  \BibitemOpen
  \bibfield  {author} {\bibinfo {author} {\bibfnamefont {C.~L.}\ \bibnamefont
  {Hammer}}\ and\ \bibinfo {author} {\bibfnamefont {L.~J.}\ \bibnamefont
  {Laslett}},\ }\bibfield  {title} {\bibinfo {title} {{Resonant Beam Extraction
  from an A. G. Synchrotron}},\ }\href {https://doi.org/10.1063/1.1717299}
  {\bibfield  {journal} {\bibinfo  {journal} {Rev. Sci. Instr.}\ }\textbf
  {\bibinfo {volume} {32}},\ \bibinfo {pages} {144} (\bibinfo {year}
  {2004})}\BibitemShut {NoStop}%
\bibitem [{\citenamefont {Kobayashi}\ and\ \citenamefont
  {Takahashi}(1967)}]{kobayashi:1967}%
  \BibitemOpen
  \bibfield  {author} {\bibinfo {author} {\bibfnamefont {Y.}~\bibnamefont
  {Kobayashi}}\ and\ \bibinfo {author} {\bibfnamefont {H.}~\bibnamefont
  {Takahashi}},\ }\bibfield  {title} {\bibinfo {title} {Improvement of the
  emittance in the resonant ejection},\ }in\ \href@noop {} {\emph {\bibinfo
  {booktitle} {{Proc. HEACC 1967}}}},\ \bibinfo {editor} {edited by\ \bibinfo
  {editor} {\bibfnamefont {R.~A.}\ \bibnamefont {Mack}}},\ \bibinfo
  {organization} {Cambridge Electron Accelerator}\ (\bibinfo  {publisher}
  {Cambridge Electron Accelerator},\ \bibinfo {year} {1967})\ pp.\ \bibinfo
  {pages} {347--351}\BibitemShut {NoStop}%
\bibitem [{\citenamefont {Barton}(1971)}]{gordon:1971}%
  \BibitemOpen
  \bibfield  {author} {\bibinfo {author} {\bibfnamefont {M.~Q.}\ \bibnamefont
  {Barton}},\ }\bibfield  {title} {\bibinfo {title} {Beam extraction from
  synchrotrons},\ }in\ \href {{https://lss.fnal.gov/conf/C710920/p85.pdf}}
  {\emph {\bibinfo {booktitle} {{Proc. HEACC 1971}}}},\ \bibinfo {editor}
  {edited by\ \bibinfo {editor} {\bibfnamefont {M.~H.}\ \bibnamefont
  {Blewett}}\ and\ \bibinfo {editor} {\bibfnamefont {N.}~\bibnamefont
  {Vogt-Nilsen}}},\ \bibinfo {organization} {CERN}\ (\bibinfo  {publisher}
  {CERN},\ \bibinfo {year} {1971})\ pp.\ \bibinfo {pages} {85--88}\BibitemShut
  {NoStop}%
\bibitem [{\citenamefont {Badano}\ \emph {et~al.}(1999)\citenamefont {Badano},
  \citenamefont {Benedikt}, \citenamefont {Bryant}, \citenamefont {Crescenti},
  \citenamefont {Holy}, \citenamefont {Maier}, \citenamefont {Pullia},
  \citenamefont {Rossi},\ and\ \citenamefont {Knaus}}]{PIMMSvol1}%
  \BibitemOpen
  \bibfield  {author} {\bibinfo {author} {\bibfnamefont {L.}~\bibnamefont
  {Badano}}, \bibinfo {author} {\bibfnamefont {M.}~\bibnamefont {Benedikt}},
  \bibinfo {author} {\bibfnamefont {P.~J.}\ \bibnamefont {Bryant}}, \bibinfo
  {author} {\bibfnamefont {M.}~\bibnamefont {Crescenti}}, \bibinfo {author}
  {\bibfnamefont {P.}~\bibnamefont {Holy}}, \bibinfo {author} {\bibfnamefont
  {A.~T.}\ \bibnamefont {Maier}}, \bibinfo {author} {\bibfnamefont
  {M.}~\bibnamefont {Pullia}}, \bibinfo {author} {\bibfnamefont
  {S.}~\bibnamefont {Rossi}},\ and\ \bibinfo {author} {\bibfnamefont
  {P.}~\bibnamefont {Knaus}} (\bibinfo {collaboration} {CERN-TERA
  Foundation-MedAustron Oncology-2000}),\ }\href@noop {} {\emph {\bibinfo
  {title} {{Proton-Ion Medical Machine Study (PIMMS), 1}}}},\ \bibinfo {type}
  {Tech. Rep.}\ (\bibinfo  {institution} {CERN},\ \bibinfo {year}
  {1999})\BibitemShut {NoStop}%
\bibitem [{\citenamefont {Hardt}(1981)}]{Hardt:1025914}%
  \BibitemOpen
  \bibfield  {author} {\bibinfo {author} {\bibfnamefont {W.}~\bibnamefont
  {Hardt}},\ }\href {https://cds.cern.ch/record/1025914} {\emph {\bibinfo
  {title} {{Ultraslow extraction out of LEAR}}}},\ \bibinfo {type} {Tech.
  Rep.}\ (\bibinfo  {institution} {CERN},\ \bibinfo {address} {Geneva},\
  \bibinfo {year} {1981})\BibitemShut {NoStop}%
\bibitem [{\citenamefont {Pullia}(2000)}]{PulliaSX}%
  \BibitemOpen
  \bibfield  {author} {\bibinfo {author} {\bibfnamefont {M.}~\bibnamefont
  {Pullia}},\ }\href {https://cds.cern.ch/record/447688} {\emph {\bibinfo
  {title} {{Transverse aspects of the slowly extracted beam}}}},\ \bibinfo
  {type} {Tech. Rep.}\ (\bibinfo  {institution} {CERN},\ \bibinfo {address}
  {Geneva},\ \bibinfo {year} {2000})\BibitemShut {NoStop}%
\bibitem [{\citenamefont {Fraser}\ \emph
  {et~al.}(2017{\natexlab{a}})\citenamefont {Fraser}, \citenamefont
  {Spanggaard}, \citenamefont {Kadi}, \citenamefont {Borburgh}, \citenamefont
  {Bertone}, \citenamefont {Cornelis}, \citenamefont {Vincke}, \citenamefont
  {Stein}, \citenamefont {Kain}, \citenamefont {Goddard}, \citenamefont
  {Roncarolo}, \citenamefont {Bj{\"o}rkman}, \citenamefont {Balhan},
  \citenamefont {Al{\'i}a}, \citenamefont {Bartosik}, \citenamefont {Gatignon},
  \citenamefont {Conan}, \citenamefont {Velotti}, \citenamefont {Mereghetti},
  \citenamefont {Stoel},\ and\ \citenamefont {Schicho}}]{Fraser2017JACoWS}%
  \BibitemOpen
  \bibfield  {author} {\bibinfo {author} {\bibfnamefont {M.~A.}\ \bibnamefont
  {Fraser}}, \bibinfo {author} {\bibfnamefont {J.}~\bibnamefont {Spanggaard}},
  \bibinfo {author} {\bibfnamefont {Y.}~\bibnamefont {Kadi}}, \bibinfo {author}
  {\bibfnamefont {J.}~\bibnamefont {Borburgh}}, \bibinfo {author}
  {\bibfnamefont {C.}~\bibnamefont {Bertone}}, \bibinfo {author} {\bibfnamefont
  {K.}~\bibnamefont {Cornelis}}, \bibinfo {author} {\bibfnamefont
  {H.}~\bibnamefont {Vincke}}, \bibinfo {author} {\bibfnamefont
  {O.}~\bibnamefont {Stein}}, \bibinfo {author} {\bibfnamefont
  {V.}~\bibnamefont {Kain}}, \bibinfo {author} {\bibfnamefont {B.}~\bibnamefont
  {Goddard}}, \bibinfo {author} {\bibfnamefont {F.}~\bibnamefont {Roncarolo}},
  \bibinfo {author} {\bibfnamefont {D.}~\bibnamefont {Bj{\"o}rkman}}, \bibinfo
  {author} {\bibfnamefont {B.}~\bibnamefont {Balhan}}, \bibinfo {author}
  {\bibfnamefont {R.~G.}\ \bibnamefont {Al{\'i}a}}, \bibinfo {author}
  {\bibfnamefont {H.}~\bibnamefont {Bartosik}}, \bibinfo {author}
  {\bibfnamefont {L.}~\bibnamefont {Gatignon}}, \bibinfo {author}
  {\bibfnamefont {N.}~\bibnamefont {Conan}}, \bibinfo {author} {\bibfnamefont
  {F.~M.}\ \bibnamefont {Velotti}}, \bibinfo {author} {\bibfnamefont
  {A.}~\bibnamefont {Mereghetti}}, \bibinfo {author} {\bibfnamefont
  {L.}~\bibnamefont {Stoel}},\ and\ \bibinfo {author} {\bibfnamefont
  {P.}~\bibnamefont {Schicho}},\ }\bibfield  {title} {{\selectlanguage
  {english}\bibinfo {title} {{SPS Slow Extraction Losses and Activation:
  Challenges and Possibilities for Improvement}}},\ }in\ \href
  {https://doi.org/10.18429/JACoW-IPAC2017-MOPIK045} {{\selectlanguage
  {english}\emph {\bibinfo {booktitle} {Proc. IPAC'17}}}}\ (\bibinfo
  {publisher} {JACoW Publishing, Geneva, Switzerland},\ \bibinfo {year}
  {2017})\ pp.\ \bibinfo {pages} {611--614}\BibitemShut {NoStop}%
\bibitem [{\citenamefont {Fraser}\ \emph {et~al.}(2019)\citenamefont {Fraser},
  \citenamefont {Goddard}, \citenamefont {Kain}, \citenamefont {Pari},
  \citenamefont {Velotti}, \citenamefont {Stoel},\ and\ \citenamefont
  {Benedikt}}]{PhysRevAccelBeams.22.123501}%
  \BibitemOpen
  \bibfield  {author} {\bibinfo {author} {\bibfnamefont {M.~A.}\ \bibnamefont
  {Fraser}}, \bibinfo {author} {\bibfnamefont {B.}~\bibnamefont {Goddard}},
  \bibinfo {author} {\bibfnamefont {V.}~\bibnamefont {Kain}}, \bibinfo {author}
  {\bibfnamefont {M.}~\bibnamefont {Pari}}, \bibinfo {author} {\bibfnamefont
  {F.~M.}\ \bibnamefont {Velotti}}, \bibinfo {author} {\bibfnamefont {L.~S.}\
  \bibnamefont {Stoel}},\ and\ \bibinfo {author} {\bibfnamefont
  {M.}~\bibnamefont {Benedikt}},\ }\bibfield  {title} {\bibinfo {title}
  {{Demonstration of slow extraction loss reduction with the application of
  octupoles at the CERN Super Proton Synchrotron}},\ }\href
  {https://doi.org/10.1103/PhysRevAccelBeams.22.123501} {\bibfield  {journal}
  {\bibinfo  {journal} {Phys. Rev. Accel. Beams}\ }\textbf {\bibinfo {volume}
  {22}},\ \bibinfo {pages} {123501} (\bibinfo {year} {2019})}\BibitemShut
  {NoStop}%
\bibitem [{\citenamefont {Nagaslaev}\ \emph {et~al.}(2019)\citenamefont
  {Nagaslaev}, \citenamefont {Brown},\ and\ \citenamefont
  {Tomizawa}}]{PhysRevAccelBeams.22.043501}%
  \BibitemOpen
  \bibfield  {author} {\bibinfo {author} {\bibfnamefont {V.}~\bibnamefont
  {Nagaslaev}}, \bibinfo {author} {\bibfnamefont {K.~A.}\ \bibnamefont
  {Brown}},\ and\ \bibinfo {author} {\bibfnamefont {M.}~\bibnamefont
  {Tomizawa}},\ }\bibfield  {title} {\bibinfo {title} {Third integer resonance
  extraction with presence of higher multipoles},\ }\href
  {https://doi.org/10.1103/PhysRevAccelBeams.22.043501} {\bibfield  {journal}
  {\bibinfo  {journal} {Phys. Rev. Accel. Beams}\ }\textbf {\bibinfo {volume}
  {22}},\ \bibinfo {pages} {043501} (\bibinfo {year} {2019})}\BibitemShut
  {NoStop}%
\bibitem [{\citenamefont {Fraser}\ \emph
  {et~al.}(2017{\natexlab{b}})\citenamefont {Fraser} \emph
  {et~al.}}]{Fraser:IPAC2017-MOPIK048}%
  \BibitemOpen
  \bibfield  {author} {\bibinfo {author} {\bibfnamefont {M.}~\bibnamefont
  {Fraser}} \emph {et~al.},\ }\bibfield  {title} {{\selectlanguage
  {english}\bibinfo {title} {{E}xperimental {R}esults of {C}rystal{-A}ssisted
  {S}low {E}xtraction at the {SPS}}},\ }in\ \href
  {https://doi.org/https://doi.org/10.18429/JACoW-IPAC2017-MOPIK048}
  {{\selectlanguage {english}\emph {\bibinfo {booktitle} {Proc. of
  International Particle Accelerator Conference (IPAC'17), Copenhagen, Denmark,
  14-19 May, 2017}}}},\ \bibinfo {series and number} {International Particle
  Accelerator Conference}\ (\bibinfo  {publisher} {JACoW},\ \bibinfo {address}
  {Geneva, Switzerland},\ \bibinfo {year} {2017})\ pp.\ \bibinfo {pages}
  {623--626},\ \bibinfo {note}
  {https://doi.org/10.18429/JACoW-IPAC2017-MOPIK048}\BibitemShut {NoStop}%
\bibitem [{\citenamefont {Giacomelli}\ \emph {et~al.}(2025)\citenamefont
  {Giacomelli}, \citenamefont {Sota}, \citenamefont {Dutheil}, \citenamefont
  {Fraser}, \citenamefont {Gorn},\ and\ \citenamefont
  {Velotti}}]{giacomelli:ipac25-tupb027}%
  \BibitemOpen
  \bibfield  {author} {\bibinfo {author} {\bibfnamefont {M.~P.}\ \bibnamefont
  {Giacomelli}}, \bibinfo {author} {\bibfnamefont {P.~A.~A.}\ \bibnamefont
  {Sota}}, \bibinfo {author} {\bibfnamefont {Y.}~\bibnamefont {Dutheil}},
  \bibinfo {author} {\bibfnamefont {M.}~\bibnamefont {Fraser}}, \bibinfo
  {author} {\bibfnamefont {A.}~\bibnamefont {Gorn}},\ and\ \bibinfo {author}
  {\bibfnamefont {F.}~\bibnamefont {Velotti}},\ }\bibfield  {title}
  {{\selectlanguage {english}\bibinfo {title} {{Improving the SPS beam
  extraction efficiency by implementation of a crystal septum}}},\ }in\ \href
  {https://doi.org/10.18429/JACoW-IPAC2025-TUPB027} {{\selectlanguage
  {english}\emph {\bibinfo {booktitle} {Proc. IPAC'25}}}},\ \bibinfo {series
  and number} {\bibinfo {series} {International Particle Accelerator
  Conference}\ No.~\bibinfo {number} {16}}\ (\bibinfo  {publisher} {JACoW
  Publishing, Geneva, Switzerland},\ \bibinfo {year} {2025})\ pp.\ \bibinfo
  {pages} {1021--1024}\BibitemShut {NoStop}%
\bibitem [{\citenamefont {Velotti}\ \emph {et~al.}(2017)\citenamefont
  {Velotti}, \citenamefont {Fraser}, \citenamefont {Goddard}, \citenamefont
  {Kain}, \citenamefont {Scandale},\ and\ \citenamefont
  {Stoel}}]{velotti:ipac17-mopik050}%
  \BibitemOpen
  \bibfield  {author} {\bibinfo {author} {\bibfnamefont {F.~M.}\ \bibnamefont
  {Velotti}}, \bibinfo {author} {\bibfnamefont {M.~A.}\ \bibnamefont {Fraser}},
  \bibinfo {author} {\bibfnamefont {B.}~\bibnamefont {Goddard}}, \bibinfo
  {author} {\bibfnamefont {V.}~\bibnamefont {Kain}}, \bibinfo {author}
  {\bibfnamefont {W.}~\bibnamefont {Scandale}},\ and\ \bibinfo {author}
  {\bibfnamefont {L.~S.}\ \bibnamefont {Stoel}},\ }\bibfield  {title}
  {{\selectlanguage {english}\bibinfo {title} {{Reduction of Resonant Slow
  Extraction Losses with Shadowing of Septum Wires by a Bent Crystal}}},\ }in\
  \href {https://doi.org/10.18429/JACoW-IPAC2017-MOPIK050} {{\selectlanguage
  {english}\emph {\bibinfo {booktitle} {Proc. IPAC'17}}}}\ (\bibinfo
  {publisher} {JACoW Publishing, Geneva, Switzerland},\ \bibinfo {year}
  {2017})\ pp.\ \bibinfo {pages} {631--634}\BibitemShut {NoStop}%
\bibitem [{\citenamefont {Esposito}\ \emph {et~al.}(2019)\citenamefont
  {Esposito}, \citenamefont {Addesa}, \citenamefont {Afonin}, \citenamefont
  {Bestman}, \citenamefont {Borg}, \citenamefont {Butcher}, \citenamefont
  {Calviani}, \citenamefont {Chesnokov}, \citenamefont {Denisov}, \citenamefont
  {Di~Castro}, \citenamefont {Donz\'e}, \citenamefont {Durum}, \citenamefont
  {Fraser}, \citenamefont {Galluccio}, \citenamefont {Garattini}, \citenamefont
  {Gavrikov}, \citenamefont {Gilardoni}, \citenamefont {Goddard}, \citenamefont
  {Hall}, \citenamefont {Iacoangeli}, \citenamefont {Ivanov}, \citenamefont
  {James}, \citenamefont {Kain}, \citenamefont {Kovalenko}, \citenamefont
  {Koznov}, \citenamefont {Lendaro}, \citenamefont {Maisheev}, \citenamefont
  {Malyarenko}, \citenamefont {Masi}, \citenamefont {Murtas}, \citenamefont
  {Natochii}, \citenamefont {Pari}, \citenamefont {Pesaresi}, \citenamefont
  {Prieto}, \citenamefont {Rossi}, \citenamefont {Sandomirskiy}, \citenamefont
  {Scandale}, \citenamefont {Seidenbinder}, \citenamefont {Serrano~Galvez},
  \citenamefont {Skorobogatov}, \citenamefont {Stoel}, \citenamefont {Taratin},
  \citenamefont {Velotti}, \citenamefont {Yanovich},\ and\ \citenamefont
  {Zhovkovska}}]{esposito:ipac19-wepmp028}%
  \BibitemOpen
  \bibfield  {author} {\bibinfo {author} {\bibfnamefont {L.~S.}\ \bibnamefont
  {Esposito}}, \bibinfo {author} {\bibfnamefont {F.}~\bibnamefont {Addesa}},
  \bibinfo {author} {\bibfnamefont {A.}~\bibnamefont {Afonin}}, \bibinfo
  {author} {\bibfnamefont {P.}~\bibnamefont {Bestman}}, \bibinfo {author}
  {\bibfnamefont {J.}~\bibnamefont {Borg}}, \bibinfo {author} {\bibfnamefont
  {M.}~\bibnamefont {Butcher}}, \bibinfo {author} {\bibfnamefont
  {M.}~\bibnamefont {Calviani}}, \bibinfo {author} {\bibfnamefont
  {Y.}~\bibnamefont {Chesnokov}}, \bibinfo {author} {\bibfnamefont
  {A.}~\bibnamefont {Denisov}}, \bibinfo {author} {\bibfnamefont
  {M.}~\bibnamefont {Di~Castro}}, \bibinfo {author} {\bibfnamefont
  {M.}~\bibnamefont {Donz\'e}}, \bibinfo {author} {\bibfnamefont
  {A.}~\bibnamefont {Durum}}, \bibinfo {author} {\bibfnamefont
  {M.}~\bibnamefont {Fraser}}, \bibinfo {author} {\bibfnamefont
  {F.}~\bibnamefont {Galluccio}}, \bibinfo {author} {\bibfnamefont
  {M.}~\bibnamefont {Garattini}}, \bibinfo {author} {\bibfnamefont
  {Y.}~\bibnamefont {Gavrikov}}, \bibinfo {author} {\bibfnamefont
  {S.}~\bibnamefont {Gilardoni}}, \bibinfo {author} {\bibfnamefont
  {B.}~\bibnamefont {Goddard}}, \bibinfo {author} {\bibfnamefont
  {G.}~\bibnamefont {Hall}}, \bibinfo {author} {\bibfnamefont {F.}~\bibnamefont
  {Iacoangeli}}, \bibinfo {author} {\bibfnamefont {Y.}~\bibnamefont {Ivanov}},
  \bibinfo {author} {\bibfnamefont {T.}~\bibnamefont {James}}, \bibinfo
  {author} {\bibfnamefont {V.}~\bibnamefont {Kain}}, \bibinfo {author}
  {\bibfnamefont {A.}~\bibnamefont {Kovalenko}}, \bibinfo {author}
  {\bibfnamefont {M.~A.}\ \bibnamefont {Koznov}}, \bibinfo {author}
  {\bibfnamefont {J.}~\bibnamefont {Lendaro}}, \bibinfo {author} {\bibfnamefont
  {V.}~\bibnamefont {Maisheev}}, \bibinfo {author} {\bibfnamefont
  {L.}~\bibnamefont {Malyarenko}}, \bibinfo {author} {\bibfnamefont
  {A.}~\bibnamefont {Masi}}, \bibinfo {author} {\bibfnamefont {F.}~\bibnamefont
  {Murtas}}, \bibinfo {author} {\bibfnamefont {A.}~\bibnamefont {Natochii}},
  \bibinfo {author} {\bibfnamefont {M.}~\bibnamefont {Pari}}, \bibinfo {author}
  {\bibfnamefont {M.}~\bibnamefont {Pesaresi}}, \bibinfo {author}
  {\bibfnamefont {J.}~\bibnamefont {Prieto}}, \bibinfo {author} {\bibfnamefont
  {R.}~\bibnamefont {Rossi}}, \bibinfo {author} {\bibfnamefont
  {Y.}~\bibnamefont {Sandomirskiy}}, \bibinfo {author} {\bibfnamefont
  {W.}~\bibnamefont {Scandale}}, \bibinfo {author} {\bibfnamefont
  {R.}~\bibnamefont {Seidenbinder}}, \bibinfo {author} {\bibfnamefont
  {P.}~\bibnamefont {Serrano~Galvez}}, \bibinfo {author} {\bibfnamefont
  {V.}~\bibnamefont {Skorobogatov}}, \bibinfo {author} {\bibfnamefont
  {L.}~\bibnamefont {Stoel}}, \bibinfo {author} {\bibfnamefont
  {A.}~\bibnamefont {Taratin}}, \bibinfo {author} {\bibfnamefont {F.~M.}\
  \bibnamefont {Velotti}}, \bibinfo {author} {\bibfnamefont {A.}~\bibnamefont
  {Yanovich}},\ and\ \bibinfo {author} {\bibfnamefont {V.}~\bibnamefont
  {Zhovkovska}},\ }\bibfield  {title} {{\selectlanguage {english}\bibinfo
  {title} {{Crystal for Slow Extraction Loss-Reduction of the SPS Electrostatic
  Septum}}},\ }in\ \href {https://doi.org/10.18429/JACoW-IPAC2019-WEPMP028}
  {{\selectlanguage {english}\emph {\bibinfo {booktitle} {Proc. IPAC'19}}}}\
  (\bibinfo  {publisher} {JACoW Publishing, Geneva, Switzerland},\ \bibinfo
  {year} {2019})\ pp.\ \bibinfo {pages} {2379--2382}\BibitemShut {NoStop}%
\bibitem [{\citenamefont {Velotti}\ \emph
  {et~al.}(2019{\natexlab{a}})\citenamefont {Velotti}, \citenamefont {Addesa},
  \citenamefont {Afonin}, \citenamefont {Bestmann}, \citenamefont {Borg},
  \citenamefont {Butcher}, \citenamefont {Calviani}, \citenamefont {Chesnokov},
  \citenamefont {Denisov}, \citenamefont {Castro}, \citenamefont {Donz\'e},
  \citenamefont {Durum}, \citenamefont {Esposito}, \citenamefont {Fraser},
  \citenamefont {Galluccio}, \citenamefont {Garattini}, \citenamefont
  {Gavrikov}, \citenamefont {Gilardoni}, \citenamefont {Goddard}, \citenamefont
  {Hall}, \citenamefont {Iacoangeli}, \citenamefont {Ivanov}, \citenamefont
  {James}, \citenamefont {Kain}, \citenamefont {Koznov}, \citenamefont
  {Lendaro}, \citenamefont {Maisheev}, \citenamefont {Malyarenko},
  \citenamefont {Masi}, \citenamefont {Mirarchi}, \citenamefont {Murtas},
  \citenamefont {Pari}, \citenamefont {Pesaresi}, \citenamefont {Prieto},
  \citenamefont {Redaelli}, \citenamefont {Rossi}, \citenamefont
  {Sandomirskiy}, \citenamefont {Scandale}, \citenamefont {Seidenbinder},
  \citenamefont {Galvez}, \citenamefont {Skorobogatov}, \citenamefont {Stoel},
  \citenamefont {Yanovich}, \citenamefont {Zamantzas},\ and\ \citenamefont
  {Zhovkovska}}]{velotti:ipac19-thxxplm2}%
  \BibitemOpen
  \bibfield  {author} {\bibinfo {author} {\bibfnamefont {F.~M.}\ \bibnamefont
  {Velotti}}, \bibinfo {author} {\bibfnamefont {F.}~\bibnamefont {Addesa}},
  \bibinfo {author} {\bibfnamefont {A.}~\bibnamefont {Afonin}}, \bibinfo
  {author} {\bibfnamefont {P.}~\bibnamefont {Bestmann}}, \bibinfo {author}
  {\bibfnamefont {J.}~\bibnamefont {Borg}}, \bibinfo {author} {\bibfnamefont
  {M.}~\bibnamefont {Butcher}}, \bibinfo {author} {\bibfnamefont
  {M.}~\bibnamefont {Calviani}}, \bibinfo {author} {\bibfnamefont
  {Y.}~\bibnamefont {Chesnokov}}, \bibinfo {author} {\bibfnamefont
  {A.}~\bibnamefont {Denisov}}, \bibinfo {author} {\bibfnamefont {M.~D.}\
  \bibnamefont {Castro}}, \bibinfo {author} {\bibfnamefont {M.}~\bibnamefont
  {Donz\'e}}, \bibinfo {author} {\bibfnamefont {A.}~\bibnamefont {Durum}},
  \bibinfo {author} {\bibfnamefont {L.~S.}\ \bibnamefont {Esposito}}, \bibinfo
  {author} {\bibfnamefont {M.}~\bibnamefont {Fraser}}, \bibinfo {author}
  {\bibfnamefont {F.}~\bibnamefont {Galluccio}}, \bibinfo {author}
  {\bibfnamefont {M.}~\bibnamefont {Garattini}}, \bibinfo {author}
  {\bibfnamefont {Y.}~\bibnamefont {Gavrikov}}, \bibinfo {author}
  {\bibfnamefont {S.}~\bibnamefont {Gilardoni}}, \bibinfo {author}
  {\bibfnamefont {B.}~\bibnamefont {Goddard}}, \bibinfo {author} {\bibfnamefont
  {G.}~\bibnamefont {Hall}}, \bibinfo {author} {\bibfnamefont {F.}~\bibnamefont
  {Iacoangeli}}, \bibinfo {author} {\bibfnamefont {Y.}~\bibnamefont {Ivanov}},
  \bibinfo {author} {\bibfnamefont {T.}~\bibnamefont {James}}, \bibinfo
  {author} {\bibfnamefont {V.}~\bibnamefont {Kain}}, \bibinfo {author}
  {\bibfnamefont {M.~A.}\ \bibnamefont {Koznov}}, \bibinfo {author}
  {\bibfnamefont {J.}~\bibnamefont {Lendaro}}, \bibinfo {author} {\bibfnamefont
  {V.}~\bibnamefont {Maisheev}}, \bibinfo {author} {\bibfnamefont
  {L.}~\bibnamefont {Malyarenko}}, \bibinfo {author} {\bibfnamefont
  {A.}~\bibnamefont {Masi}}, \bibinfo {author} {\bibfnamefont {D.}~\bibnamefont
  {Mirarchi}}, \bibinfo {author} {\bibfnamefont {F.}~\bibnamefont {Murtas}},
  \bibinfo {author} {\bibfnamefont {M.}~\bibnamefont {Pari}}, \bibinfo {author}
  {\bibfnamefont {M.}~\bibnamefont {Pesaresi}}, \bibinfo {author}
  {\bibfnamefont {J.}~\bibnamefont {Prieto}}, \bibinfo {author} {\bibfnamefont
  {S.}~\bibnamefont {Redaelli}}, \bibinfo {author} {\bibfnamefont
  {R.}~\bibnamefont {Rossi}}, \bibinfo {author} {\bibfnamefont
  {Y.}~\bibnamefont {Sandomirskiy}}, \bibinfo {author} {\bibfnamefont
  {W.}~\bibnamefont {Scandale}}, \bibinfo {author} {\bibfnamefont
  {R.}~\bibnamefont {Seidenbinder}}, \bibinfo {author} {\bibfnamefont {P.~S.}\
  \bibnamefont {Galvez}}, \bibinfo {author} {\bibfnamefont {V.}~\bibnamefont
  {Skorobogatov}}, \bibinfo {author} {\bibfnamefont {L.}~\bibnamefont {Stoel}},
  \bibinfo {author} {\bibfnamefont {A.}~\bibnamefont {Yanovich}}, \bibinfo
  {author} {\bibfnamefont {C.}~\bibnamefont {Zamantzas}},\ and\ \bibinfo
  {author} {\bibfnamefont {V.}~\bibnamefont {Zhovkovska}},\ }\bibfield  {title}
  {{\selectlanguage {english}\bibinfo {title} {{Demonstration of Loss Reduction
  Using a Thin Bent Crystal to Shadow an Electrostatic Septum During Resonant
  Slow Extraction}}},\ }in\ \href
  {https://doi.org/10.18429/JACoW-IPAC2019-THXXPLM2} {{\selectlanguage
  {english}\emph {\bibinfo {booktitle} {Proc. IPAC'19}}}}\ (\bibinfo
  {publisher} {JACoW Publishing, Geneva, Switzerland},\ \bibinfo {year}
  {2019})\ pp.\ \bibinfo {pages} {3399--3403}\BibitemShut {NoStop}%
\bibitem [{\citenamefont {Velotti}\ \emph
  {et~al.}(2019{\natexlab{b}})\citenamefont {Velotti}, \citenamefont
  {Esposito}, \citenamefont {Fraser}, \citenamefont {Kain}, \citenamefont
  {Gilardoni}, \citenamefont {Goddard}, \citenamefont {Pari}, \citenamefont
  {Prieto}, \citenamefont {Rossi}, \citenamefont {Scandale}, \citenamefont
  {Stoel}, \citenamefont {Galluccio}, \citenamefont {Garattini},\ and\
  \citenamefont {Gavrikov}}]{PhysRevAccelBeams.22.093502}%
  \BibitemOpen
  \bibfield  {author} {\bibinfo {author} {\bibfnamefont {F.~M.}\ \bibnamefont
  {Velotti}}, \bibinfo {author} {\bibfnamefont {L.~S.}\ \bibnamefont
  {Esposito}}, \bibinfo {author} {\bibfnamefont {M.~A.}\ \bibnamefont
  {Fraser}}, \bibinfo {author} {\bibfnamefont {V.}~\bibnamefont {Kain}},
  \bibinfo {author} {\bibfnamefont {S.}~\bibnamefont {Gilardoni}}, \bibinfo
  {author} {\bibfnamefont {B.}~\bibnamefont {Goddard}}, \bibinfo {author}
  {\bibfnamefont {M.}~\bibnamefont {Pari}}, \bibinfo {author} {\bibfnamefont
  {J.}~\bibnamefont {Prieto}}, \bibinfo {author} {\bibfnamefont
  {R.}~\bibnamefont {Rossi}}, \bibinfo {author} {\bibfnamefont
  {W.}~\bibnamefont {Scandale}}, \bibinfo {author} {\bibfnamefont {L.~S.}\
  \bibnamefont {Stoel}}, \bibinfo {author} {\bibfnamefont {F.}~\bibnamefont
  {Galluccio}}, \bibinfo {author} {\bibfnamefont {M.}~\bibnamefont
  {Garattini}},\ and\ \bibinfo {author} {\bibfnamefont {Y.}~\bibnamefont
  {Gavrikov}},\ }\bibfield  {title} {\bibinfo {title} {Septum shadowing by
  means of a bent crystal to reduce slow extraction beam loss},\ }\href
  {https://doi.org/10.1103/PhysRevAccelBeams.22.093502} {\bibfield  {journal}
  {\bibinfo  {journal} {Phys. Rev. Accel. Beams}\ }\textbf {\bibinfo {volume}
  {22}},\ \bibinfo {pages} {093502} (\bibinfo {year}
  {2019}{\natexlab{b}})}\BibitemShut {NoStop}%
\bibitem [{\citenamefont {Velotti}\ \emph {et~al.}(2022)\citenamefont
  {Velotti}, \citenamefont {Di~Castro}, \citenamefont {Esposito}, \citenamefont
  {Fraser}, \citenamefont {Gilardoni}, \citenamefont {Goddard}, \citenamefont
  {Kain},\ and\ \citenamefont {Matheson}}]{velotti:ipac22-wepost013}%
  \BibitemOpen
  \bibfield  {author} {\bibinfo {author} {\bibfnamefont {F.~M.}\ \bibnamefont
  {Velotti}}, \bibinfo {author} {\bibfnamefont {M.}~\bibnamefont {Di~Castro}},
  \bibinfo {author} {\bibfnamefont {L.~S.}\ \bibnamefont {Esposito}}, \bibinfo
  {author} {\bibfnamefont {M.}~\bibnamefont {Fraser}}, \bibinfo {author}
  {\bibfnamefont {S.}~\bibnamefont {Gilardoni}}, \bibinfo {author}
  {\bibfnamefont {B.}~\bibnamefont {Goddard}}, \bibinfo {author} {\bibfnamefont
  {V.}~\bibnamefont {Kain}},\ and\ \bibinfo {author} {\bibfnamefont
  {E.}~\bibnamefont {Matheson}},\ }\bibfield  {title} {{\selectlanguage
  {english}\bibinfo {title} {{Exploitation of Crystal Shadowing via
  Multi-Crystal Array, Optimisers and Reinforcement Learning}}},\ }in\ \href
  {https://doi.org/10.18429/JACoW-IPAC2022-WEPOST013} {{\selectlanguage
  {english}\emph {\bibinfo {booktitle} {Proc. IPAC'22}}}},\ \bibinfo {series
  and number} {\bibinfo {series} {International Particle Accelerator
  Conference}\ No.~\bibinfo {number} {13}}\ (\bibinfo  {publisher} {JACoW
  Publishing, Geneva, Switzerland},\ \bibinfo {year} {2022})\ pp.\ \bibinfo
  {pages} {1707--1710}\BibitemShut {NoStop}%
\bibitem [{\citenamefont {Veres}\ \emph
  {et~al.}(2025{\natexlab{b}})\citenamefont {Veres}, \citenamefont {Bartosik},
  \citenamefont {Franchetti}, \citenamefont {Giovannozzi},\ and\ \citenamefont
  {Paraschou}}]{DEV_IPAC25_SPS_LPR}%
  \BibitemOpen
  \bibfield  {author} {\bibinfo {author} {\bibfnamefont {D.~E.}\ \bibnamefont
  {Veres}}, \bibinfo {author} {\bibfnamefont {H.}~\bibnamefont {Bartosik}},
  \bibinfo {author} {\bibfnamefont {G.}~\bibnamefont {Franchetti}}, \bibinfo
  {author} {\bibfnamefont {M.}~\bibnamefont {Giovannozzi}},\ and\ \bibinfo
  {author} {\bibfnamefont {K.}~\bibnamefont {Paraschou}},\ }\bibfield  {title}
  {\bibinfo {title} {{Demonstrating Beam Splitting Through Stable Islands
  Formed by the Third-Order Resonance at the CERN Super Proton Synchrotron}},\
  }\href {https://doi.org/10.1088/1742-6596/3094/1/012037} {\bibfield
  {journal} {\bibinfo  {journal} {J. Phys. Conf. Ser.}\ }\textbf {\bibinfo
  {volume} {3094}},\ \bibinfo {pages} {012037} (\bibinfo {year}
  {2025}{\natexlab{b}})}\BibitemShut {NoStop}%
\bibitem [{Note1()}]{Note1}%
  \BibitemOpen
  \bibinfo {note} {Note that the normalized nonlinear gradient is defined as
  $k_n=1/(B \rho ) \dd ^n B_y/\dd x^n$, where $B \rho $ is the rigidity of the
  beam.}\BibitemShut {Stop}%
\bibitem [{\citenamefont {Kain}\ \emph {et~al.}(2019)\citenamefont {Kain},
  \citenamefont {Velotti}, \citenamefont {Fraser}, \citenamefont {Goddard},
  \citenamefont {Prieto}, \citenamefont {Stoel},\ and\ \citenamefont
  {Pari}}]{sx_COSE}%
  \BibitemOpen
  \bibfield  {author} {\bibinfo {author} {\bibfnamefont {V.}~\bibnamefont
  {Kain}}, \bibinfo {author} {\bibfnamefont {F.~M.}\ \bibnamefont {Velotti}},
  \bibinfo {author} {\bibfnamefont {M.~A.}\ \bibnamefont {Fraser}}, \bibinfo
  {author} {\bibfnamefont {B.}~\bibnamefont {Goddard}}, \bibinfo {author}
  {\bibfnamefont {J.}~\bibnamefont {Prieto}}, \bibinfo {author} {\bibfnamefont
  {L.~S.}\ \bibnamefont {Stoel}},\ and\ \bibinfo {author} {\bibfnamefont
  {M.}~\bibnamefont {Pari}},\ }\bibfield  {title} {\bibinfo {title} {Resonant
  slow extraction with constant optics for improved separatrix control at the
  extraction septum},\ }\href
  {https://doi.org/10.1103/PhysRevAccelBeams.22.101001} {\bibfield  {journal}
  {\bibinfo  {journal} {Phys. Rev. Accel. Beams}\ }\textbf {\bibinfo {volume}
  {22}},\ \bibinfo {pages} {101001} (\bibinfo {year} {2019})}\BibitemShut
  {NoStop}%
\end{thebibliography}%
\end{document}